# Rerouting shortest paths in planar graphs

Paul Bonsma

Computer Science Department, Humboldt University Berlin,
Unter den Linden 6, 10099 Berlin. `bonsma@informatik.hu-berlin.de`

**Abstract.** A rerouting sequence is a sequence of shortest $st$-paths such that consecutive paths differ in one vertex. We study the the Shortest Path Rerouting Problem, which asks, given two shortest $st$-paths $P$ and $Q$ in a graph $G$, whether a rerouting sequence exists from $P$ to $Q$. This problem is PSPACE-hard in general, but we show that it can be solved in polynomial time if $G$ is planar. To this end, we introduce a dynamic programming method for reconfiguration problems.

## 1 Introduction

In recent literature, various *reconfiguration problems* have been studied [11]. For example, Gopalan et al [9] studied the following problem. Given a boolean formula $\phi$, the *solution graph* $G(\phi)$ has as vertex set all *solutions*, i.e. variable assignments that satisfy $\phi$. Two solutions are adjacent if they differ in exactly one variable. An instance of the *Satisfiability Reconfiguration* problem consists a boolean formula $\phi$, and two solutions $s$ and $t$ for $\phi$. The question is whether in $G(\phi)$, a path from $s$ to $t$ exists. In [9], boolean formulas are classified into those based on *tight relations* and *non-tight relations*. It is shown that in the tight case, the Satisfiability Reconfiguration problem can be solved in polynomial time, and in the non-tight case, the problem is PSPACE-complete. A similar dichotomy result is proved in [9] for the problem of deciding whether $G(\phi)$ is connected.

Similar results have been obtained for reconfiguration problems based on other combinatorial problems, such as vertex coloring [3,4,5,6], independent set [10,11,14], and more [8,11,12,13]. To obtain reconfiguration versions of these problems, one only needs to define a (symmetric) adjacency relation between solutions. One of the motivations for this research is to study the structure of the solution space of well-studied combinatorial problems, which can explain the performance of various heuristics [4,9]. In addition, similar problems have also occurred in practical applications such as stacking problems in storage spaces [16] and train switchyards (see [15] and references therein). Most of these problems turned out to be PSPACE-complete on general instances.

Various deep hardness results have been proved for reconfiguration problems, such as the first two (independent) PSPACE-hardness results for such problems, on satisfiability reconfiguration [9] and sliding block puzzles [10]. To our knowledge, the other known PSPACE-hardness results on reconfiguration problems have been proved using reductions from these two results. We remark that various PSPACE-complete problems of a similar flavor have been described earlier, such as in the context of local search [17]. An essential difference is however that these are based on asymmetric adjacency relations.

In contrast, very few deep (or even non-trivial) algorithmic results have been obtained for reconfiguration problems. (One of the few exceptions is the result by Cereceda et al [6] on the reconfiguration of vertex colorings using three colors.) A reason for this lack may be that no general algorithmic techniques are known. These two problems impede progress in this area, and might erroneously suggest that reconfiguration problems are only interesting from a complexity theoretic viewpoint. Addressing these problems is the main motivation for the research presented in this paper.

In this paper, we study the *Shortest Path Rerouting (SPR)* problem, as introduced by Kamiński et al [13]. Given is a graph $G$, with vertices $s, t \in V(G)$, and two shortest $st$-paths $P$ and $Q$. Shortest paths are adjacent if they differ in one vertex. The question is whether there exists a *rerouting sequence* from $P$ to $Q$, which is a sequence of shortest $st$-paths $Q_0, \ldots, Q_k$ with $Q_0 = P$, $Q_k = Q$, such that consecutive paths are adjacent.

In [13], instances are described where the minimum length of a rerouting sequence is exponential in $n = |V(G)|$. Secondly, it is shown that that it is strongly NP-hard to decide whether a rerouting sequence of length at most $k$ exists. In [2], SPR is proved to be PSPACE-complete, and polynomial time algorithms are given for the case where $G$ is claw-free or chordal.

Our main result is that SPR can be decided in polynomial time for planar graphs. Since (rerouting) shortest paths is an important concept in networking, and many networks in practice are planar, this is a relevant graph class for this problem. To obtain this result, we develop a dynamic programming method for reconfiguration problems. Our dynamic programming for SPR returns the correct answer for all instances, but may require more than polynomial time for some. Nevertheless, in appendix D we show that this algorithm is not only useful for deciding planar SPR: in the case where every vertex that lies on a shortest $st$-path has at most two neighbors closer to $s$, and at most two neighbors further from $s$, this dynamic programming algorithm decides SPR in polynomial time as well. Secondly, our results illustrate the importance and strength of searching for *central solutions* in the solution graph. These algorithmic techniques are discussed in a broader context in Section 6. In the next section, we first give detailed definitions, and then give an outline of the rest of the paper. Because of space constraints, this paper only contains sketches of proofs. Statements for which detailed proofs can be found in the appendix are marked with a star.

## 2 Preliminaries

For graph theoretical notions not defined here, and background on results mentioned in this section, we refer to [7]. Walks, paths and cycles in graphs are defined as sequences of vertices. By $N(v)$ we denote the neighborhood of a vertex $v$. A *plane graph* is a graph together with an embedding in the plane (without crossing edges). For planar graphs, an embedding can be found in polynomial time, so it suffices to prove our results for plane graphs. If a plane graph is 2-connected, then the boundary of every face is a cycle, which is called a *facial cycle*. Cycles in plane graphs correspond to simple closed curves in the plane, which divide the plane into two regions. For a cycle $C$ in a plane graph $G$ and vertices $s, t \in V(G)$, we say $C$ *separates $s$ from $t$* if $s$ and $t$ lie in different regions of the curve given by $C$ (and thus $s, t \notin V(C)$). Instead of $S \cup \{x\}$ and $S \setminus \{x\}$, we write $S + x$ and $S - x$, respectively.

Throughout this paper, $s$ and $t$ denote two (distinct) vertices of an unweighted, undirected, simple, finite graph $G$, and we will consider shortest $st$-paths in $G$. Let $d$ denote the distance from $s$ to $t$ in $G$. For $i \in \{0, \ldots, d\}$, we define $L_i \subseteq V(G)$ to be the set of vertices that lie on a shortest $st$-path, at distance $i$ from $s$. The vertex set $L_i$ is also called a *layer*.

Observe that a shortest $st$-path $P$ contains exactly one vertex of every layer, and that shortest paths are uniquely determined by their vertex set. Therefore, we will denote shortest $st$-paths $P$ by their *vertex set*. A vertex set $Q \subseteq V(G)$ such that there exists a shortest $st$-path $P$ with $Q \subseteq P$ is called a *shortest $st$-subpath*. Since we are only concerned with shortest paths in $G$ between two given terminals $s$ and $t$, we will call these *S-paths* for short. Shortest $st$-subpaths will be called *S-subpaths*. When considering reduced instances, defined by the subgraph induced by all shortest paths between two vertices $x$ and $y$, these definitions refer to $x$ and $y$. A rerouting step from an S-path $P$ to $Q$ consists of replacing a vertex $a \in P$ by another vertex $b$ in the same layer, such that an S-path results. To be precise, let $x, a, y \in P$ such that $x \in L_{i-1}$, $a \in L_i$ and $y \in L_{i+1}$. For any $b \in L_i$ with $\{x, y\} \subseteq N(b)$, the *rerouting step $x, a, y \to x, b, y$* may be applied, which yields $Q = P - a + b$.

Let $G$ be a graph and $s, t \in V(G)$. The *rerouting graph* $\mathrm{SP}(G, s, t)$ has as set of vertices all S-paths in $G$. Two paths are adjacent if they differ in exactly one vertex. To distinguish vertices of $\mathrm{SP}(G, s, t)$ from vertices of $G$, the former will be called *nodes*. Subsets $S \subseteq V(\mathrm{SP}(G, s, t))$ will be called sets of S-paths or sets of nodes, depending on the context. In order to prove our results, we need to consider two additional variants of the SPR problem, which are defined by considering different adjacency relations. Call a rerouting step $x, a, y \to x, b, y$ a *restricted rerouting step* if $ab \in E(G)$. In the *restricted rerouting graph* $\mathrm{SP}^R(G, s, t)$, two S-paths $P$ and $Q$ are adjacent if $Q$ can be obtained from $P$ using a restricted rerouting step.

In the case that $G$ is a plane graph, we can define a third type of rerouting graph. A sequence of four vertices $x, a, b, y$ is called a *switch* if $x, a, y, b, x$ is a cycle that separates $s$ from $t$, and for some $i$, $x \in L_i$ and $y \in L_{i+2}$. The vertices $x$ and $y$ are called its *(left and right) switch vertices*. Together, they are also called a *switch-pair*. For instance, the graph $G_4$ shown in Figure 2 contains exactly one switch: $6, 7, 8, 9$. ($6, 8, 7, 9$ is considered to be the same switch.) For readability, in our figures some edges are shown as pairs of half edges in our figures; edges leaving the top of the figure continue on the bottom. A rerouting step $x, a, y \to x, b, y$ is called *topological* if $x, a, b, y$ is not a switch. In the *topological rerouting graph* $\mathrm{SP}^T(G, s, t)$, two S-paths $P$ and $Q$ are adjacent if $Q$ can be obtained from $P$ by a topological rerouting step.

Walks in $\mathrm{SP}(G, s, t)$, $\mathrm{SP}^R(G, s, t)$ and $\mathrm{SP}^T(G, s, t)$ are called *rerouting sequences*, *restricted rerouting sequences*, and *topological rerouting sequences* respectively. Let $P$ be an S-path, and let $Q$ be an S-subpath. We write $P \rightsquigarrow_G Q$ to denote that in $G$, there exists a rerouting sequence from $P$ to an S-path $Q'$ with $Q \subseteq Q'$. Similarly, the relations $P \rightsquigarrow_G^R Q$ and $P \rightsquigarrow_G^T Q$ are used for the restricted and topological case, respectively. If the graph $G$ in question is clear, the subscript is omitted. If $P \rightsquigarrow Q$, $P \rightsquigarrow^R Q$ or $P \rightsquigarrow^T Q$, we also say that $Q$ is *reachable* from $P$. We write $P \not\rightsquigarrow Q$ to denote that $P \rightsquigarrow Q$ does not hold.

The *Generalized Shortest Path Rerouting (GSPR) Problem* asks, given a graph $G$ with $s, t \in V(G)$, an S-path $P$ and an S-subpath $Q$, whether $P \rightsquigarrow_G Q$. Similarly, for the *Restricted SPR (RSPR) Problem* and *Topological SPR (TSPR) Problem*, it should be decided whether $P \rightsquigarrow_G^R Q$ and $P \rightsquigarrow_G^T Q$, respectively.

Since $Q$ may be an *S-subpath* (in all of these problems), GSPR is a generalization of the SPR problem (defined in Section 1). The RSPR Problem in turn generalizes the GSPR Problem: a GSPR instance $G, P, Q$ can easily be transformed to an equivalent RSPR instance by adding edges between every pair of vertices in the same layer. (This may destroy planarity however.)

All algorithmic results and reductions presented in this paper for deciding $P \rightsquigarrow Q$, $P \rightsquigarrow^T Q$ or $P \rightsquigarrow^R Q$ are *constructive*, in the following sense: if $P \rightsquigarrow Q$ for an S-subpath $Q$, then in addition an S-path $Q'$ with $P \rightsquigarrow Q'$ and $Q \subseteq Q'$ can be constructed with the same complexity. (Analog for the topological and restricted case.) For brevity, this is not stated in every lemma and theorem, but this fact is essential for the proofs in Section 5.

**Outline** In Section 3, we present a dynamic programming algorithm for the RSPR Problem. This algorithm returns the correct answer for all instances, but may require exponential time in some cases. In Section 4 we show however that for instances in a certain standard form, the algorithm always terminates in polynomial time. This is used to prove that TSPR can be solved in polynomial time, by giving a transformation to RSPR instances in standard form. Finally, in Section 5 we show how to handle switches in planar graphs, and give an algorithm for SPR in planar graphs. This algorithm reduces any instance of the problem to a polynomial number of instances of the TSPR Problem.

## 3 A Dynamic Programming Algorithm for RSPR

Let $P$ be an S-path in $G$, and $Q$ be an S-subpath. We want to decide whether $P \rightsquigarrow_G^R Q$. For $i = 1, \ldots, d-1$, we define the graph $G_i$ as follows: $G_i$ is obtained from $G$ by first removing all vertices in layers $L_{i+1}, \ldots, L_{d-1}$, and then adding edges from $t$ to all vertices in $L_i$. By $P^i$ and $Q^i$ we denote $P \cap V(G_i)$ and $Q \cap V(G_i)$, which clearly are again an S-path and an S-subpath in $G_i$. Figure 1(a) shows an example of $G_3$, $P^3$ and $Q^3$ (for instance, for the $G$, $P$ and $Q$ that are shown in Figure 4(a), although Figure 4(a) contains no layer edges).

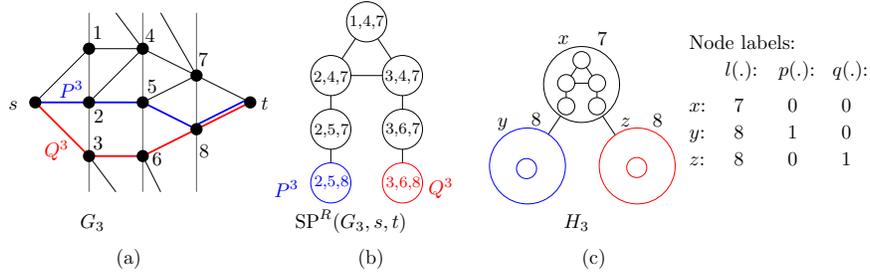

**Fig. 1.** $G_3$, its restricted rerouting graph $\mathrm{SP}^R(G_3, s, t)$, and a contraction $H_3$ of it (with nodes $x, y, z$), which is the encoding of $G_3, P, Q$. In the nodes of $\mathrm{SP}^R(G_3, s, t)$, the vertices of the corresponding paths are shown, except $s$ and $t$. In the nodes of $H_3$, the corresponding contracted subgraph of $\mathrm{SP}^R(G_3, s, t)$ is drawn.

The idea is now to compute $\mathrm{SP}^R(G_{i+1}, s, t)$ from $\mathrm{SP}^R(G_i, s, t)$, for $i = 0, \ldots, d-2$. In the end, this will yield $\mathrm{SP}^R(G_{d-1}, s, t) = \mathrm{SP}^R(G, s, t)$, and we can decide whether in this graph a path from $P$ to $Q$ exists. The problem is of course that the graphs $\mathrm{SP}^R(G_i, s, t)$ may be exponentially large compared to $G$. We solve this problem by instead considering a graph $H_i$ that is obtained from a component of $\mathrm{SP}^R(G_i, s, t)$ by contracting connected subgraphs into single nodes, and using node labels to keep track of essential information about the corresponding path sets.

For two S-paths $R$ and $R'$ in $G_i$, we define $R \sim_i R'$ if and only if there exists a restricted rerouting sequence from $R$ to $R'$ that does not change the vertex in $L_i$ (so $R \cap L_i = R' \cap L_i$). Clearly, $\sim_i$ is an equivalence relation. Furthermore, if $S$ is an equivalence class of $\sim_i$, then $S$ induces a connected subgraph of $\mathrm{SP}^R(G_i, s, t)$. These are exactly the subgraphs of $\mathrm{SP}^R(G_i, s, t)$ that we will contract to obtain $H_i$. The following definition is illustrated in Figure 1(b) and (c).

**Definition 1 (Encoding)** *Let $P$ be an S-path in $G$ of length $d$, let $Q$ be an S-subpath in $G$, and let $i \in \{0, \ldots, d-1\}$. The encoding $H_i$ of $G_i, P, Q$ is obtained from $H' = SP^R(G_i, s, t)$ as follows:*

1. *Delete every component of $H'$ that does not contain the node $P^i$.*
2. *For every equivalence class $S \subseteq V(H')$ of $\sim_i$ that has not been deleted, contract the subgraph $H'[S]$ into a single node $x$, and define $S_x := S$ (this is a set of S-paths in $G_i$). Define $l(x)$ to be the vertex in $L_i$ that is part of every path in $S_x$. Set $p(x) = 1$ if $P^i \in S_x$, and $p(x) = 0$ otherwise. Set $q(x) = 1$ if there exists an S-path $Q' \in S_x$ with $Q^i \subseteq Q'$ and $q(x) = 0$ otherwise.*

Note that the encoding $H_i$ defined this way is unique. We first observe that this definition serves the main purpose of allowing us to decide whether $P \leadsto^R Q$.

**Proposition 2 (*)** *Let $H_{d-1}$ be the encoding of $G_{d-1}, P, Q$. Then $P \leadsto^R Q$ if and only if $H_{d-1}$ contains a node $x$ with $q(x) = 1$.*

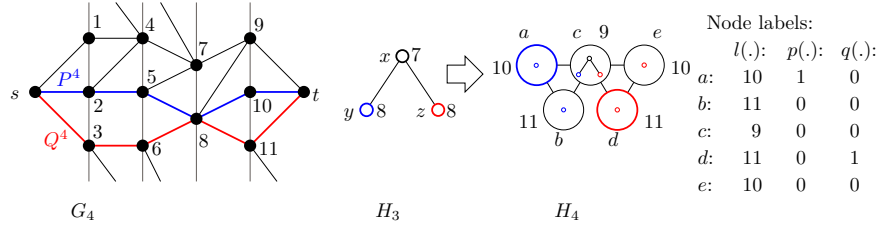

**Fig. 2.** Constructing the encoding $H_4$ from $H_3$. In every node $a$ of $H_4$, the corresponding subgraph $C_a$ of $H_3$ is drawn. Numbers next to nodes $a$ indicate their label $l(a)$.

Next, we study how the encoding $H_{i+1}$ of $G_{i+1}, P, Q$ is related to the encoding $H_i$ of $G_i, P, Q$. The objective is that we wish to construct $H_{i+1}$ from $H_i$ without considering $\mathrm{SP}^R(G_{i+1}, s, t)$. An example of this construction is shown in Figure 2. For every $v \in L_{i+1}$, let $X_v$ be the set of nodes of $H_i$ that correspond to neighbors of $v$. Formally, $X_v := \{x \in V(H_i) \mid l(x) \in N(v)\}$. By $H_i^v := H_i[X_v]$ we denote the subgraph of $H_i$ induced by these nodes. This graph may have multiple components, even though $H_i$ is connected. For $v \in L_{i+1}$ and $x \in V(H_i)$, by $S_x \oplus v$ we denote the set obtained by adding $v$ to every path in $S_x$, so $S_x \oplus v = \cup_{R \in S_x}(R+v)$. We will use this notation for the case where $v$ is adjacent to $l(x)$, so then $S_x \oplus v$ is a set of S-paths in $G_{i+1}$.

**Lemma 3 (*)** *Let $H_i$ and $H_{i+1}$ be the encodings of $G_i, P, Q$ and $G_{i+1}, P, Q$, respectively. For any $v \in L_{i+1}$ and component $C$ of $H_i^v$, $\cup_{x \in V(C)}(S_x \oplus v)$ is a set of S-paths in $G_{i+1}$ that is an equivalence class of $\sim_{i+1}$. In addition, for every $a \in V(H_{i+1})$ with $l(a) = v$, there exists a component $C_a$ of $H_i^v$ such that $S_a = \cup_{x \in V(C_a)}(S_x \oplus v)$.*

Lemma 3 shows that for every $a \in V(H_{i+1})$, there exists a corresponding component $C$ of $H_i^v$, where $v = l(a)$. We denote this component by $C_a$.

**Lemma 4 (*)** *Let $H_{i+1}$ be the encoding of $G_{i+1}, P, Q$. Let $a, b \in V(H_{i+1})$.*
  *(i) $ab \in E(H_{i+1})$ if and only if $l(a)l(b) \in E(G)$ and $V(C_a) \cap V(C_b) \neq \emptyset$.*
  *(ii) $p(a) = 1$ if and only if $l(a) \in P$ and there exists a node $x \in V(C_a)$ with $p(x) = 1$.*
  *(iii) $q(a) = 1$ if and only if $Q \cap L_{i+1} \subseteq \{l(a)\}$ and there exists a node $x \in V(C_a)$ with $q(x) = 1$.*

The previous two lemmas give all the information that is necessary to compute $H_{i+1}$ from $H_i$, including the node labels $l$, $p$ and $q$. The essential fact that will guarantee a good complexity bound is that for this computation, knowledge of the path sets $S_x$ for $x \in V(H_i)$ is unnecessary. This yields a dynamic programming algorithm for deciding $P \rightsquigarrow^R Q$ (Proposition 2).

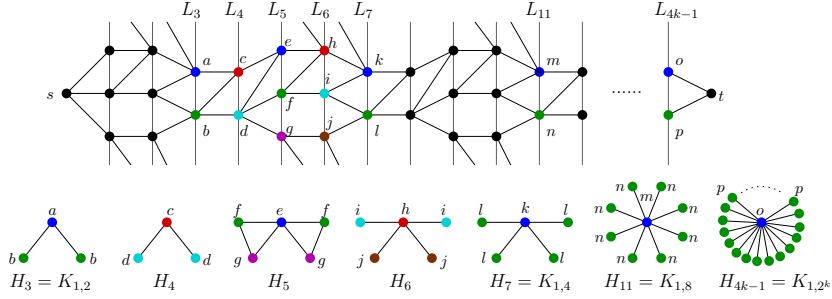

**Fig. 3.** A plane instance of RSPR where the encoding $H_i$ becomes exponentially large. Colors and numbers next to nodes $x \in V(H_i)$ indicate the labels $l(x)$.

**Theorem 5 (*)** *Let $G$ be a graph on $n$ vertices with two vertices $s,t \in V(G)$ at distance $d$. Let $P$ be an S-path, and let $Q$ be an S-subpath in $G$. In time polynomial in $n$ and $m$, it can be decided whether $P \rightsquigarrow^R Q$. Here $m = \max_{i \in \{1,\ldots,d-1\}} |V(H_i)|$, where $H_i$ is the encoding of $G_i, P, Q$.*

Theorem 5 shows that the RSPR problem can be solved in polynomial time if the size of the encodings $H_i$ remains polynomially bounded. However, since the problem is PSPACE-hard [2], we should not expect this to be true for all graphs. Indeed, there exist examples where the size of the encoding grows exponentially. The example shown in Figure 3 shows that this even true for the case of planar graphs of maximum degree 6 (4 when ignoring edges not on S-paths). It can be verified that for $i = 4k - 1$, the encoding $H_i$ is a star with $2^k$ leaves.

However, there are many nontrivial instances for which this algorithm is polynomial. For instance, we remark (without proof) that this holds for the class of instances described by Kamiński et al [13], where any rerouting sequence from $P$ to $Q$ has exponential length. (Provided that $s$ and $t$ are swapped.) Next, we prove that $H_i$ remains polynomially bounded for instances in a certain standard form, which is closely related to planar graphs. In Appendix D, we show that this also holds for instances where every vertex has at most two neighbors in both the previous and the next layer. Figure 3 shows that this result is sharp; the condition cannot be replaced by maximum degree 4.

## 4 A Polynomial Complexity Bound for TSPR

In this section, we show that if $G, P, Q$ is a (reduced) TSPR instance, then in polynomial time it can be decided whether $P \rightsquigarrow^T Q$. To this end, we define a standard form for RSPR instances, and show for these that the algorithm from Section 3 terminates in polynomial time. Subsequently we show how TSPR instances can be transformed to equivalent RSPR instances in standard form. Figure 4(b) illustrates the following definition.

**Definition 6** *Consider a graph $G$ and vertices $s, t \in V(G)$ that are part of an RSPR instance. Then $G$ is in* standard form *if the following properties hold:*

(i) For every $i \in \{1, \ldots, d-1\}$, $G[L_i]$ has maximum degree 2.
(ii) For every $i \in \{1, \ldots, d-1\}$ and $v \in L_i$, $G[L_{i-1} \cap N(v)]$ is a path.
(iii) For every $i$ and $u, v \in L_i$, if $uv \in E(G)$ then $|N(u) \cap N(v) \cap L_{i-1}| \leq 1$.

A *homomorphism* from a (simple) graph $H$ to a (simple) graph $G$ is a function $\phi : V(H) \to V(G)$ such that for all $uv \in E(H)$, $\phi(u)\phi(v) \in E(G)$. Such a homomorphism is *locally injective* if for every $u \in V(H)$ and $v, w \in N(u)$, $\phi(v) \neq \phi(w)$.

**Theorem 7 (*)** *Let $G, P, Q$ be an RSPR instance in standard form. Then in polynomial time, it can be decided whether $P \leadsto^R Q$.*

**Proof sketch:** Theorem 5 shows that it suffices to show that for every $i$, the size of encoding $H_i$ of $G_i, P, Q$ remains polynomially bounded in $|V(G)|$. By Lemma 4(i), $l$ is a homomorphism from $H_i$ to $G[L_i]$, for every $i$. Using Properties (ii) and (iii) of Definition 6, it can be proved by induction over $i$ that $l$ is in fact locally injective. Then, using Property (i), it follows that $H_i$ has maximum degree at most 2 as well, so it is a path or a cycle.

For $v \in L_i$, we denote by $n_i(v)$ the number of nodes $x \in V(H_i)$ with $l(x) = v$. Define $\text{MAX}_i := \max_{v \in L_i} n_i(v)$. Since $H_i$ is a path or a cycle, and the labels $l$ form a locally injective homomorphism to the path or cycle $G[L_i]$, it can be shown that for any $u \in L_{i+1}$ and any component $C_a$ of $H_i^u$, all vertices in $N(u) \cap L_i$ occur as labels in $C_a$, unless $H_i$ is a path and $C_a$ contains one of its end vertices. There are at most two end vertices, so $\text{MAX}_{i+1} \leq \text{MAX}_i + 2$ follows. An easy induction proof then shows that for all $i \in \{1, \ldots, d-1\}$, $|V(H_i)| \leq 2i|L_i|$. □

The objective is to apply the above result for TSPR instances $G, P, Q$. To this end, we will give a polynomial transformation from to an equivalent instance $G', P, Q$ of RSPR, and prove that the latter instance is in standard form. In the case of TSPR and GSPR, it will be useful to work with reduced instances. An instance $G, P, Q$ of GSPR or TSPR is *reduced* if:
1. Every vertex and edge of $G$ lies on an S-path,
2. $G$ contains no cut vertices, and
3. $G$ contains no *neighborhood-dominated vertices*, which are vertices $z$ for which there exists a vertex $z'$ with $N(z) \subseteq N(z')$.

We remark that, even though the above definition is useful for both GSPR and TSPR, only in the case of GSPR we can give a polynomial time procedure that can transform every instance to a set of equivalent reduced instances. This procedure is straightforward: we may simply delete all vertices and edges not on S-paths. As long as there exists a neighborhood-dominated vertex $z$, we may delete $z$, and replace occurrences of $z$ in $P$ and $Q$ by the corresponding vertex $z'$. When a cut vertex $v$ is present, the instance basically consists of two independent instances: one induced by all shortest $sv$-paths, and one induced by all shortest $vt$-paths.

**Theorem 8 (*)** *Let $G, P, Q$ be a GSPR instance. In polynomial time, a set of reduced GSPR instances can be constructed such that $P \leadsto_G Q$ if and only if for* every reduced instance $G_i, P_i, Q_i$, it holds that $P_i \rightsquigarrow_{G_i} Q_i$. If $G$ is plane, all of the reduced instances are plane. The sum of the number of edges of the reduced instances is at most $|E(G)|$.

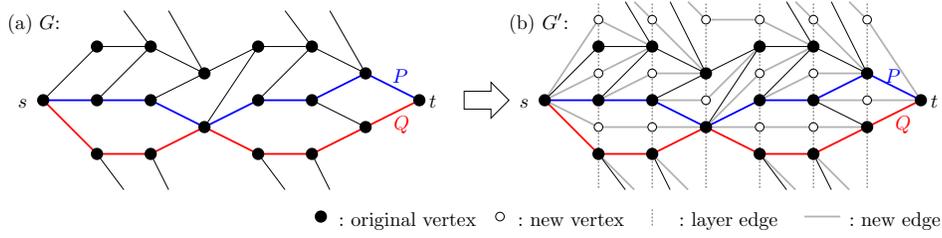

**Fig. 4.** The transformation of a TSPR instance $G, P, Q$ to an equivalent RSPR instance $G', P, Q$.

For $v \in V(G)$, by $\mathrm{dist}_s(v)$ we denote the distance from $s$ to $v$, so $v \in L_{\mathrm{dist}_s(v)}$. We assume that $G, P, Q$ is reduced, so every vertex and edge of $G$ lies on an S-path. It follows that for every edge $uv$, it holds that $|\mathrm{dist}_s(u) - \mathrm{dist}_s(v)| = 1$. Furthermore, $G$ is 2-connected. So for every face $f$ of $G$ and vertex $v$ incident with $f$, $v$ has exactly two incident edges $uv$ and $vw$ that are also incident with $f$. We call $v$ a *local maximum for $f$* if $\mathrm{dist}_s(v) > \mathrm{dist}_s(u) = \mathrm{dist}_s(w)$, and a *local minimum for $f$* if $\mathrm{dist}_s(v) < \mathrm{dist}_s(u) = \mathrm{dist}_s(w)$.

**Proposition 9 (*)** *Let $G$ be a 2-connected plane graph in which every vertex and edge lies on a shortest st-path. For every face $f$ of $G$, there is exactly one local maximum and one local minimum.*

Now we can define the transformation from the TSPR instance $G$ to the RSPR instance $G'$. This transformation is illustrated in Figure 4, and consists of the following two steps.

1. For every face $f$ of $G$, we do the following. Let $u$ and $v$ be the local minimum and maximum of $f$. Since $G$ is simple, $\mathrm{dist}_s(u) \leq \mathrm{dist}_s(v) - 2$. Let $\ell = \mathrm{dist}_s(v) - \mathrm{dist}_s(u)$. Add $\ell - 1$ *new vertices* $x_1, \ldots, x_{\ell-1}$, drawn in the face $f$, and $\ell$ edges such that $u, x_1, \ldots, x_{\ell-1}, v$ is a path of length $\ell$, drawn in face $f$. Clearly, this preserves planarity. Call the vertices and edges introduced this way *new vertices and edges*. The vertices and edges that were already present in $G$ are called *original vertices and edges*.
2. For every face $f$ in the resulting graph, we do the following. Note that $f$ still has a unique local minimum $u$ and local maximum $v$. Furthermore, for every edge $xy$, $|\mathrm{dist}_s(x) - \mathrm{dist}_s(y)| = 1$. It follows that for every $i \in \{\mathrm{dist}_s(u)+1, \ldots, \mathrm{dist}_s(v)-1\}$, there are exactly two vertices $a$ and $b$ incident with $f$ in layer $L_i$. Between every such pair of vertices $a$ and $b$, we can add an edge $ab$, drawn in the face $f$, without destroying planarity. Call the new edges *layer edges*.

Call the resulting plane graph $G'$. It can be shown that $G', P, Q$ is an RSPR instance in standard form. Since $G$ contains no neighborhood-dominated vertices, for any topological rerouting step $x, a, y \to x, b, y$, it holds that $x, a, y, b, x$ is a facial cycle. So it can be replaced by two restricted rerouting steps $x, a, y \to x, z, y \to x, b, y$ for $G'$, where $z$ is a new vertex. Therefore, $P \leadsto_G^T Q$ implies $P \leadsto_{G'}^R Q$. For the converse, it can be shown that rerouting steps in a restricted rerouting sequence for $G'$ can be grouped in pairs $x, a, y \to x, z, y \to x, b, y$ where only $z$ is a new vertex. Hence $x, a, y, b, x$ is a facial cycle of $G$, and $x, a, y \to x, b, y$ is a topological rerouting step. We conclude that $P \leadsto_G^T Q$ if and only if $P \leadsto_{G'}^R Q$. The above transformation is polynomial, so applying Theorem 7 gives:

**Theorem 10 (*)** *Let $G, P, Q$ be a reduced TSPR instance. In polynomial time, it can be decided whether $P \leadsto_G^T Q$.*

## 5 Switches, general rerouting and the main theorem

As shown in Figure 3, the presence of switches in a plane graph $G$ may cause our dynamic programming algorithm to take exponential time. However, in this section we show that switches also give a lot of structural information, which can be used to obtain a polynomial time algorithm. If $x, a, b, y$ is a switch in $G$, then in many cases we can reduce the problem to two smaller subproblems, defined as follows: $G_{sy}$ is the subgraph of $G$ induced by all vertices that lie on a shortest $sy$-path, and $G_{xt}$ is the subgraph of $G$ induced by all vertices that lie on a shortest $xt$-path. For an S-subpath $Q$, we denote $Q_{xt} = Q \cap V(G_{xt})$, and $Q_{sy} = Q \cap V(G_{sy})$. We remark that the rerouting sequences that we consider in $G_{sy}$ ($G_{xt}$), consist of shortest $sy$-paths (resp. $xt$-paths). We are now ready to state the key lemma for reducing the GSPR problem, when switches are present.

**Lemma 11 (*)** *Let $G, P, Q$ be a plane reduced GSPR instance, such that $x, y$ is a switch pair with $\{x, y\} \subseteq P$, and $Q$ is one of the following:*
 *(i) an S-path that contains $x$ and $y$,*
 *(ii) $Q = \{x', y'\}$ where $x', y'$ is a switch pair, or*
 *(iii) $|Q| = 1$.*
*Then $P \leadsto_G Q$ if and only if both $P_{sy} \leadsto_{G_{sy}} Q_{sy}$ and $P_{xt} \leadsto_{G_{xt}} Q_{xt}$.*

**Theorem 12** *Let $G, P, Q$ be a plane reduced GSPR instance, where $Q$ is a set containing a switch pair or a single vertex of $G$. Then in polynomial time it can be decided whether $P \leadsto Q$.*

**Proof:** First, compute whether $P \leadsto^T Q$, which can be done in polynomial time (Theorem 10). If yes, then clearly $P \leadsto Q$ holds as well, and an S-path $Q'$ with $P \leadsto Q'$ and $Q \subseteq Q'$ can be computed. (Recall that, as discussed in Section 2, all of our algorithms are constructive.) If $P \not\leadsto^T Q$, then for every switch pair $x, y$, we compute whether $P \leadsto^T \{x, y\}$, and if so, compute the corresponding reachable S-path that contains $x$ and $y$. Since the number of switch pairs in $G$ is polynomial (linear in fact), this can again be done in polynomial time (Theorem 10). If no switch pair is reachable, then we may conclude that $P \not\leadsto Q$.

Now consider a switch pair $x, y$ with $P \leadsto^T \{x, y\}$. Let $P'$ be the S-path with $P \leadsto^T P'$ and $\{x, y\} \subseteq P'$ that has been computed. Clearly, $P \leadsto^T P'$ implies $P \rightsquigarrow P'$ and $P' \rightsquigarrow P$. Therefore, $P \rightsquigarrow Q$ if and only if $P' \rightsquigarrow Q$. By Lemma 11, $P' \rightsquigarrow Q$ if and only if both $P'_{xt} \rightsquigarrow Q_{xt}$ and $P'_{sy} \rightsquigarrow Q_{sy}$ hold. We decide the latter two properties recursively. This way, we can decide whether $P \rightsquigarrow Q$.

It remains to consider the complexity of this algorithm. We argued that the complexity of the above procedure, not counting the recursive calls, can be bounded by a (monotone increasing) polynomial poly$(n)$, where $n = |V(G)|$. Recall that $d$ denotes the distance between the end vertices $s$ and $t$. If there are no switch pairs (which is true in particular when $d \leq 3$), then the entire procedure terminates in time poly$(n)$.

For $d \geq 3$, we prove by induction over $d$ that the complexity of the algorithm can be bounded by $(\frac{4}{3}d - 3) \cdot \text{poly}(n)$. We have just proved the induction basis ($d = 3$), so now assume $d \geq 4$. In that case, the algorithm may consider a switch pair $x, y$, and reduce the problem to two instances $G_{sy}$ and $G_{xt}$. Then the distance between the end vertices of these instances is $d'$ and $d - d' + 2$, respectively, for some $3 \leq d' \leq d - 1$. Using the induction assumption and the fact that both $G_{sy}$ and $G_{xt}$ contain at most $n$ vertices, we can bound the total complexity by $(\frac{4}{3}d' - 3) \cdot \text{poly}(n) + (\frac{4}{3}d - \frac{4}{3}d' - 1) \cdot \text{poly}(n) + \text{poly}(n) = (\frac{4}{3}d - 3) \cdot \text{poly}(n)$. □

Finally, we are able to prove the main result of this paper.

**Theorem 13 (*)** *Let $G$ be a plane graph, and let $P$ and $Q$ be S-paths in $G$. In polynomial time, it can be decided whether $P \rightsquigarrow Q$.*

**Proof sketch:** By Theorem 8, we may assume that the instance is reduced. The proof is similar to the proof of Theorem 12. The difference is that now, for every switch pair $x, y$, we decide whether $P \rightsquigarrow \{x, y\}$ and $Q \rightsquigarrow \{x, y\}$, which can be done in polynomial time (Theorem 12). If the answer differs for $P$ and $Q$, then we may conclude $P \not\rightsquigarrow Q$. If both $P \rightsquigarrow \{x, y\}$ and $Q \rightsquigarrow \{x, y\}$, then the problem can be reduced using Lemma 11(i), and decided recursively. If $P \not\rightsquigarrow \{x, y\}$ and $Q \not\rightsquigarrow \{x, y\}$ for *every* switch pair $x, y$, then it suffices to decide whether $P \rightsquigarrow^T Q$ (Theorem 10). □

## 6 Discussion

In Section 3, we introduced a dynamic programming method for reconfiguration problems, that can informally be summarized as follows: identify subgraphs $G_i$, that are separated from the rest of the graph by small separators $L_i$ (in our case, the distance layers). For deciding the reconfiguration problem, detailed information about all solutions is not necessary. So based on how solutions intersect with $L_i$, one can identify large connected subgraphs (equivalence classes) in the solution graph of $G_i$, which may be contracted. This method can be applied to all kinds of reconfiguration problems, and different sets of separators, in particular small separators given by *tree decompositions* [1]. The challenge is however proving that the resulting encodings stay polynomially bounded. In our case, exponentially large star-like structures (Figure 3) could be avoided by restricting to the topological version of the problem. There may however be different approaches, such as based on reduction rules for the encodings.

Exploring this method seems useful for answering the following interesting open question: *Are there reconfiguration problems that are PSPACE-hard for graphs of bounded tree width? Or are there PSPACE-hard reconfiguration problems that can be solved in polynomial time for graphs of treewidth $k$, for every fixed $k$?* In contrast to the abundance of positive results on NP-complete problems for bounded treewidth [1], to our knowledge, not a single positive or negative result is known for reconfiguration problems on bounded treewidth graphs!

Furthermore, our proofs in Section 5 demonstrated the following simple but strong technique: Instead of trying to reconfigure $P$ directly to $Q$, one should try to reconfigure both $P$ and $Q$ simultaneously to a 'common central solution' in the solution graph $\text{SP}(G, s, t)$. Central solutions for this problem turn out to be paths that contain many switch pairs. Once both $P$ and $Q$ are reconfigured to paths that contain a maximum number of switch pairs, deciding whether $P \leadsto Q$ amounts to simply checking topological reachability. This paradigm, based on identifying central solutions in the solution graph, seems very useful for many reconfiguration problems.

## A  Proofs omitted from Section 3

Throughout this section, we will use $u$, $v$ and $w$ to refer to vertices in $G$, $x$, $y$ and $z$ to refer to nodes of $H_i$, and $a$, $b$ and $c$ to refer to nodes of $H_{i+1}$.

**Proposition 2** *Let $H_{d-1}$ be the encoding of $G_{d-1}, P, Q$. Then $P \leadsto^R Q$ if and only if $H_{d-1}$ contains a node $x$ with $q(x) = 1$.*

**Proof:** Note that $G_{d-1} = G$, $P^{d-1} = P$ and $Q^{d-1} = Q$. If $P \leadsto^R Q$ then in $\mathrm{SP}^R(G_{d-1}, s, t)$, the component that contains $P$ also contains an S-path $Q'$ with $Q \subseteq Q'$. This is the component that is contracted into $H_{d-1}$, so $H_{d-1}$ contains a node $x$ with $Q' \in S_x$, and thus $q(x) = 1$.

On the other hand, if $H_{d-1}$ contains a node $x$ with $q(x) = 1$, then $S_x$ contains an S-path $Q'$ with $Q \subseteq Q'$. Since $H_{d-1}$ is obtained by contracting the component of $\mathrm{SP}^R(G_{d-1}, s, t)$ that contains $P$, $\mathrm{SP}^R(G_{d-1}, s, t)$ has a component that contains both $P$ and $Q'$, so $P \leadsto^R Q$. □

### A.1  The proof of Lemma 3

We prove the lemma in two steps.

**Proposition 14** *Let $H_i$ be the encoding of $G_i, P, Q$, and $v \in L_{i+1}$. Let $C$ be a component of $H_i^v$. Then $\cup_{x \in V(C)}(S_x \oplus v)$ is a set of S-paths in $G_{i+1}$ that is an equivalence class of $\sim_{i+1}$.*

**Proof:** For $y \in V(C)$ and any two S-paths $R_1, R_2 \in S_y$, there exists a restricted rerouting sequence from $R_1$ to $R_2$ that uses only paths from $S_y$. Because $H_i$ is a minor of $\mathrm{SP}^R(G_i, s, t)$, for any edge $yz \in E(H_i)$, there exist paths $R_y \in S_y$ and $R_z \in S_z$ that are adjacent in $\mathrm{SP}^R(G_i, s, t)$. In particular, this holds for edges of the component $C$ as well. Since $C$ is connected, we can combine these two observations to conclude that for any two paths $R, R' \in \cup_{x \in V(C)} S_x$, there exists a restricted rerouting sequence in $\text{SP}^R(G_i, s, t)$ from $R$ to $R'$ that uses only paths in $\cup_{x \in V(C)} S_x$. Since $C$ is a component of $H_i^v$, all of these paths contain a vertex of $L_i$ that is adjacent to $v \in L_{i+1}$. So adding $v$ to all of these paths yields a restricted rerouting sequence from $R + v$ to $R' + v$ in which every path contains $v$, which proves that $\cup_{x \in V(C)}(S_x \oplus v)$ is a subset of an equivalence class of $\sim_{i+1}$.

Denote $S_a = \cup_{x \in V(C)}(S_x \oplus v)$. To conclude the proof, we show that any equivalence class of $\sim_{i+1}$ that intersects with $S_a$ is a subset of $S_a$. Consider a restricted rerouting sequence $R_0, \ldots, R_k$ (consisting of S-paths in $G_{i+1}$), that does not change the vertex $v \in L_{i+1}$. We have to show that if $R_0 \in S_a$, then $R_k \in S_a$ as well. Removing $v$ from every path in this rerouting sequence gives a sequence of paths $R'_0, \ldots, R'_k$. It suffices to show that these are all in $\cup_{x \in V(C)} S_x$. Indeed, for any $j$ and $y \in V(C)$, if $R'_j \in S_y$ and $R'_{j+1}$ is obtained from $R'_j$ by a rerouting step in a layer other than $L_i$, then by definition of $S_y$, $R'_{j+1} \in S_y$ as well. On the other hand, if $R'_{j+1}$ is obtained from $R'_j$ by a rerouting step in $L_i$, then there exists a node $z \in V(H_i)$ that is adjacent to $y$ with $R'_{j+1} \in S_z$. Furthermore, $l(z) \in N(v)$, since $R_{j+1}$ is a path. So $z$ is part of the component $C$ as well. We may conclude that all paths $R_j$ in the rerouting sequence are part of $S_a$. □

In various proofs in the appendix, we will construct sequences of S-paths such that consecutive paths either differ in one vertex, or are the same. Clearly, these can be made into rerouting sequences by removing all paths that are the same as the previous path in the sequence. In other words, by *removing duplicates*.

**Proposition 15** *Let $H_i$ and $H_{i+1}$ be the encodings of $G_i, P, Q$ and $G_{i+1}, P, Q$, respectively. For every $a \in V(H_{i+1})$, there exists a component $C_a$ of $H_i^v$ such that $S_a = \cup_{x \in V(C_a)}(S_x \oplus v)$, where $v = l(a)$.*

**Proof:** For every S-path $R \in S_a$, there exists a restricted rerouting sequence from $P^{i+1}$ to $R$, since $H_{i+1}$ is obtained by contracting the component of $\text{SP}^R(G_{i+1}, s, t)$ that contains $P^{i+1}$. Consider a restricted rerouting sequence from $P^{i+1}$ to $R$. Deleting the vertices from $L_{i+1}$ in these paths gives a restricted rerouting sequence from $P^i$ to $R - v$ in $G_i$ (after removing duplicates). So $R - v$ is part of the component of $\text{SP}^R(G_i, s, t)$ that has been contracted to obtain $H_i$, and therefore there exists a node $y \in V(H_i)$ with $R - v \in S_y$. This node $y$ is part of $H_i^v$. Let $C_a$ denote the component of $H_i^v$ that contains $y$. By Proposition 14, $\cup_{x \in V(C_a)}(S_x \oplus v)$ is an equivalence class of $\sim_{i+1}$. Considering the S-path $R$, we see that this equivalence class intersects with the equivalence class $S_a$, so they are in fact the same. □

Together, Propositions 14 and 15 give Lemma 3.

### A.2 The proof of Lemma 4

We prove the lemma in two steps. (Recall that a *minor* of a graph $G$ is a graph that can be obtained from a subgraph of $G$ by a sequence of contractions.)

**Proposition 16** Let $H_{i+1}$ be the encoding of $G_{i+1}, P, Q$. Let $a, b \in V(H_{i+1})$. Then $ab \in E(H_{i+1})$ if and only if $l(a)l(b) \in E(G)$ and $V(C_a) \cap V(C_b) \neq \emptyset$.

**Proof:** $H_{i+1}$ is a minor of $\mathrm{SP}^R(G_{i+1}, s, t)$, so $ab \in E(H_{i+1})$ if and only if there exist S-paths $R_a \in S_a$ and $R_b \in S_b$ such that $R_a$ and $R_b$ are adjacent in $\mathrm{SP}^R(G_{i+1}, s, t)$.

Suppose that such adjacent paths $R_a \in S_a$ and $R_b \in S_b$ exist. Since $H_{i+1}$ is obtained by contracting *maximal* connected subgraphs consisting of S-paths that all have the same $L_{i+1}$-vertex, it follows that $l(a) \neq l(b)$. So $R_a$ and $R_b$ differ in their $L_{i+1}$-vertex. They differ in one vertex, so $R_a - l(a) = R_b - l(b)$. This S-path $R_a - l(a)$ of $G_i$ is part of a set $S_x$ with $x \in V(C_a) \cap V(C_b)$, which therefore is nonempty. In addition, the adjacency of $R_a$ and $R_b$ in $\mathrm{SP}^R(G_{i+1}, s, t)$ implies $l(a)l(b) \in E(G)$, by definition. This proves the first direction.

On the other hand, if $V(C_a) \cap V(C_b) \neq \emptyset$ and $l(a)l(b) \in E(G)$, then we can choose a path $R \in S_x$ for some $x \in V(C_a) \cap V(C_b)$, and add the respective vertices $l(a)$ and $l(b)$ to $R$, to obtain two S-paths that are adjacent in $\mathrm{SP}^R(G_{i+1}, s, t)$. (Note that $G$ has no loops, so $l(a) \neq l(b)$.) $\square$

**Proposition 17** Let $H_{i+1}$ be the encoding of $G_{i+1}, P, Q$. Let $a \in V(H_{i+1})$.

- $p(a) = 1$ if and only if $l(a) \in P$ and there exists a node $x \in V(C_a)$ with $p(x) = 1$.
- $q(a) = 1$ if and only if $Q \cap L_{i+1} \subseteq \{l(a)\}$ and there exists a node $x \in V(C_a)$ with $q(x) = 1$.

**Proof:** We only prove the second statement; the proof of the first statement is analog. Suppose $q(a) = 1$, so $S_a$ contains an S-path $Q'$ with $Q^{i+1} \subseteq Q'$. Clearly $Q'$ contains $l(a)$. Then either $Q^{i+1}$ contains $l(a)$ as well, or $Q^{i+1} \cap L_{i+1} = \emptyset$. In addition, there exists a node $x \in C_a$ such that $S_x$ contains $Q' - l(a)$. From $Q^i \subseteq Q' - l(a)$ it follows that $q(x) = 1$. This proves the forward implication of the statement. The converse follows similarly. $\square$

Together, Propositions 16 and 17 give Lemma 4.

### A.3 The proof of Theorem 5

**Theorem 5** Let $G$ be a graph on $n$ vertices with two vertices $s, t \in V(G)$ at distance $d$. Let $P$ be an S-path, and let $Q$ be an S-subpath in $G$. In time polynomial in $n$ and $m$, it can be decided whether $P \rightsquigarrow^R Q$. Here $m = \max_{i \in \{1,\ldots,d-1\}} |V(H_i)|$, where $H_i$ is the encoding of $G_i, P, Q$.

**Proof:** The dynamic programming algorithm computes the encoding $H_i$ of $G_i, P, Q$ for $i = 0, \ldots, d - 1$, and the node labels $l(x)$, $p(x)$ and $q(x)$ for all $x \in V(H_i)$. (Note that the path sets $S_x$ are not computed.) We may initialize $H_0$ to be a graph on one node $x$, with $l(x) = s$, $p(x) = 1$ and $q(x) = 1$. To compute $H_{i+1}$ from $H_i$, we do the following:

Start with the empty graph. For every $v \in L_{i+1}$, consider the subgraph $H_i^v$ of $H_i$. For every component $C_a$ of this subgraph, introduce a node $a$ with $l(a) = v$. This way, we introduce every node that should be part of $H_{i+1}$ (Proposition 15), and possibly more. Set $p(a)$ and $q(a)$ according to Proposition 17. After introducing nodes this way for every $v \in L_{i+1}$ and every component of $H_i^v$, we add edges. Following Proposition 16, we add an edge between nodes $a$ and $b$ if and only if $l(a)l(b) \in E(G)$ and $V(C_a) \cap V(C_b) \neq \emptyset$.

It follows that this procedure yields a graph $H'$ that contains the encoding $H_{i+1}$ as subgraph. However, $H'$ is not necessarily connected, even though $H_{i+1}$ is. (Recall that the encoding $H_{i+1}$ is obtained by contracting a component of $\text{SP}^R(G_{i+1}, s, t)$.) Nevertheless, from Proposition 16 we may conclude that a node $a \in V(H')$ is part of $H_{i+1}$ if and only if it lies in the same component as the unique node $b$ with $p(b) = 1$. This shows how $H_{i+1}$ can be computed from $H_i$ for every $i$.

After constructing all of these encodings, the algorithm terminates by concluding that $P \leadsto^R Q$ if and only if $H_{d-1}$ contains a node $a$ with $q(a) = 1$, which is correct by Proposition 2.

It is easy to extend this algorithm such that it is constructive, i.e. such that it also returns an S-path $Q'$ of $G$ with $Q \subseteq Q'$, in case $P \leadsto^R Q$. This is done by storing a single *representative S-path* $Q_a$ of $G_i$ with $Q^i \subseteq Q_a$, for every $H_i$ and every $a \in V(H_i)$ with $q(a) = 1$. By Proposition 17, if $q(a) = 1$ then the component $C_a$ of $H_{i-1}$ contains at least one node $x$ with $q(x) = 1$. Adding $l(a)$ to the representative S-path $Q_x$ for $x$ then gives a representative S-path for $a$.

It remains to consider the complexity. There are $d - 1$ steps where $H_{i+1}$ is constructed, using as input $H_i$ and $G$, $P$ and $Q$, and one step where we verify whether $H_{d-1}$ contains a node $a$ with $q(a) = 1$. It is easy to see that every such step can be done in polynomial time in the input size. Therefore the total complexity can be bounded by $n^{O(1)} \cdot m^{O(1)}$, with $m = \max_{i \in \{1,\ldots,d-1\}} |V(H_i)|$.
□

## B  Proofs omitted from Section 4

### B.1  The proof of Theorem 7

A locally injective homomorphism is called a *LIH* for short. Throughout, denote by $H_i$ the encoding of $G_i, P, Q$ (Definition 1). By Proposition 16, for every $i$, $l$ is a homomorphism from $H_i$ to $G[L_i]$. Our first goal is to prove that in the case of instances in standard form, this homomorphism is locally injective.

**Lemma 18** *Let $G, P, Q$ be an RSPR instance in standard form, with distance $d$ from $s$ to $t$. For $i \in \{0, \ldots, d-1\}$, let $H_i$ be the encoding of $G_i, P, Q$. Then for every $i$, the node labels $l$ form a LIH from $H_i$ to $G[L_i]$.*

**Proof:** We prove the statement by induction over $i$. For $i = 0$, the statement is trivial. For $i = 1$, $G_i$ is a $K_k$ with $k = |L_1|$, so $H_i$ has $k$ nodes, and $l$ is a bijection to $L_i$. This proves local injectivity. Now for the induction step, suppose $l$ is a LIH from $H_i$ to $G[L_i]$, with $i \geq 1$. We prove that $l$ is a LIH from $H_{i+1}$ to $G[L_{i+1}]$.

Consider $a \in V(H_{i+1})$, and the corresponding component $C_a$ of $H_i^v$, where $v = l(a)$. As a first step, we prove that $C_a$ contains no two nodes with the same label. Suppose to the contrary that it does. Consider a shortest path $R = x_0, x_1, \ldots, x_p$ in $C_a$ between two nodes of the same label, so $l(x_0) = l(x_p)$. By the induction assumption, $l$ is a LIH from $H_i$ to $G[L_i]$, so $p \geq 3$. For $j \in \{0, \ldots, p-1\}$, define $v_j = l(x_j)$. (By definition of $H_i^v$, for every $j$, $v_j \in N(v)$ holds.) Since we chose $R$ to be a shortest path between two nodes of the same label, $v_0, \ldots, v_{p-1}$ are all distinct. Since in addition $l$ is a LIH from $H_i$ to $G[L_i]$, it follows that $v_0, \ldots, v_{p-1}$ is a path in $G[L_i]$. Finally, since $x_{p-1}$ and $x_p$ are adjacent and $l(x_p) = l(x_0) = v_0$, it follows that $v_{p-1}$ is adjacent to $v_0$. But then $v_0, \ldots, v_{p-1}, v_0$ is a cycle in $G[L_i]$, consisting of neighbors of $v$, which contradicts Definition 6(ii). We conclude that all labels in $C_a$ are distinct, for every $a \in V(H_{i+1})$.

Using this fact, we can prove that $l$ is a LIH from $H_{i+1}$ to $G[L_{i+1}]$. Suppose not, so there exists a node $a \in V(H_{i+1})$ with neighbors $b, c \in V(H_{i+1})$ such that $l(b) = l(c)$. Let $u = l(a)$ and $v = l(b) = l(c)$. This means that there exist nodes $x \in V(C_b) \cap V(C_a)$ and $y \in V(C_c) \cap V(C_a)$, and $uv \in E(G)$ (Proposition 16). By Definition 6(iii), $u$ and $v$ share at most one neighbor $w$ in the previous layer $L_i$. So $l(x) = w$ and $l(y) = w$. Because we showed above that all labels in $C_a$ are distinct, it follows that $x = y$. But this contradicts that $C_b$ and $C_c$ are distinct components of $H_i^v$. □

**Corollary 19** *Let $G, P, Q$ be an RSPR instance in standard form, with distance $d$ from $s$ to $t$. For $i \in \{1, \ldots, d-1\}$, the encoding $H_i$ of $G_i, P, Q$ is a path or a cycle.*

**Proof:** Every layer $G[L_i]$ has maximum degree at most 2 (Definition 6(i)). For every $i$, $l$ is a LIH from $H_i$ to $G[L_i]$ (Lemma 18). So $H_i$ has maximum degree at most 2 as well. By definition, encodings are connected, so $H_i$ is a path or a cycle. □

For $v \in L_i$, we denote by $n_i(v)$ the number of nodes $x \in V(H_i)$ with $l(x) = v$. Define
$$\text{MAX}_i := \max_{v \in L_i} n_i(v).$$

Using the previous lemma and corollary, we can show that $\text{MAX}_i$ grows at most linearly with $i$, which is done in the next lemma. From this it will follow that for every $i$, the size of $H_i$ is polynomially bounded.

**Lemma 20** *Let $G, P, Q$ be an RSPR instance in standard form, with distance $d$ from $s$ to $t$. For $i \in \{1, \ldots, d-2\}$, $\text{MAX}_{i+1} \leq \text{MAX}_i + 1$.*

**Proof:** The proof of this lemma is illustrated in Figure 5. We will prove the statement by showing for an arbitrary vertex $u \in L_{i+1}$ that $n_{i+1}(u) \leq \text{MAX}_i + 1$ holds. It suffices to prove that $H_i^u$ has at most $\text{MAX}_i + 1$ components (Proposition 15).

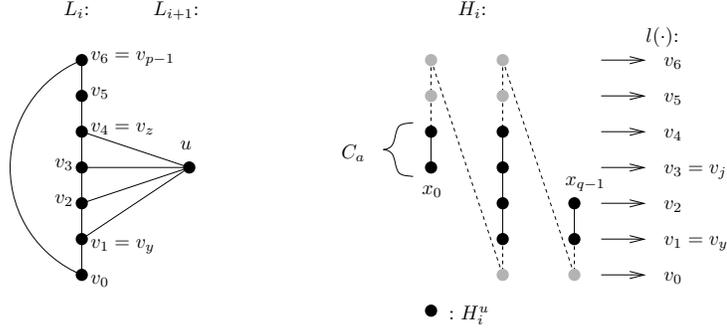

**Fig. 5.** An illustration of the proof of Lemma 20. Nodes $x$ of $H_i$ are arranged such that their height indicates their label $l(x)$. The black nodes and solid edges of $H_i$ indicate $H_i^u$. In this example, $\text{MAX}_i = 2$, but $H_i^u$ contains three components. So $H_{i+1}$ contains three nodes with label $u$, and $\text{MAX}_{i+1} = 3$.

By Definition 6(i), $G[L_i]$ has maximum degree 2, so every component is a path or a cycle. By Definition 6(ii), the neighbors of $u$ in $L_i$ induce a path, so they are all part of the same component $C$ of $G[L_i]$. Number the vertices of this component $C$ as $v_0, \ldots, v_{p-1}$, along the path or cycle. (To be precise: such that $v_0, \ldots, v_{p-1}$ is a path, or such that $v_0, \ldots, v_{p-1}, v_0$ is a cycle, respectively.) If $C$ is a cycle, then we choose these labels such that in addition, $v_{p-1} \notin N(u)$, which is possible since $G[N(u) \cap L_i]$ is a path, so at least one vertex of the cycle $C$ is not included in $N(u)$. Let $y \in \{0, \ldots, p-1\}$ be the lowest index such that $v_y \in N(u)$. Observe that $N(u) \cap L_i = \{v_y, v_{y+1}, \ldots, v_z\}$ for some $z \geq y$ (since $v_0 v_{p-1} \notin E(G[L_i \cap N(u)])$).

Since $H_i$ is a path or a cycle (Corollary 19), we may similarly number its nodes $x_0, \ldots, x_{q-1}$, such that this sequence is a path. (If $H_i$ is a cycle, then in addition $x_{q-1} x_0 \in E(H_i)$. We may also assume that $q \geq 2$, otherwise the lemma follows immediately.)

We will now show that there is at most one component $C_a$ of $H_i^u$ that contains no node with label $v_y$. Let $C_a$ be such a component. Let $j$ be the lowest index such that $C_a$ contains a node $x$ with $l(x) = v_j$. So by choice of $C_a$, $j \geq y+1$, and $x$ has no neighbor $x'$ in $H_i$ with $l(x') = v_{j-1}$. Because $l$ is a LIH from $H_i$ to $G[L_i]$ (Lemma 18) and $v_j$ has at most one neighbor other than $v_{j-1}$ (Definition 6(i)), it follows that $x$ has degree at most 1. So $H_i$ is a path, and $x$ is one of its end vertices. W.l.o.g. we may assume $x = x_0$. Using again that $l$ is a LIH, we conclude that $l(x_1) = v_{j+1 \bmod p}$, $l(x_2) = v_{j+2 \bmod p}$, etc. In particular, if $l(x_q) = v_{j'}$ (with $j' \geq 1$), then $l(x_{q-1}) = v_{j'-1}$. From these facts we may conclude that for any component $C_a$ of $H_i^u$ that contains no node with label $v_y$, it holds that $x_0 \in V(C_a)$. So there is at most one such component. It follows that $n_{i+1}(u) \leq n_i(v_y) + 1 \leq \text{MAX}_i + 1$. □

**Theorem 7** *Let $G, P, Q$ be an RSPR instance in standard form. Then in polynomial time, it can be decided whether $P \leadsto^R Q$.*

**Proof:** Let $n = |V(G)|$. We apply the dynamic programming algorithm from Section 3 to this instance to decide whether $P \leadsto^R Q$. This algorithm computes encodings $H_i$ of $G_i, P, Q$ for every $i$. A simple induction proof based on Lemma 20 shows that for every $i$, $\text{MAX}_i \leq i$, so $H_i$ contains at most $i$ nodes with the same label. Therefore, it contains at most $i \cdot |L_i| \in O(n^2)$ nodes in total. So the algorithm terminates in time $n^{O(1)}$ (Theorem 5). □

## B.2 The proof of Theorem 8

Recall that an instance $G$, $P$, $Q$ of GSPR is called *reduced* if:

1. Every vertex and edge of $G$ lies on an S-path,
2. $G$ contains no cut vertices, and
3. $G$ contains no *neighborhood-dominated vertices*, which are vertices $z$ for which there exists a vertex $z'$ with $N(z) \subseteq N(z')$.

Given an instance consisting of a (possibly plane) graph $G$, and S-path $P$ and an S-subpath $Q$, we obtain a set of reduced instances by applying the following three reduction rules exhaustively. The following proposition is trivial.

**Proposition 21 (Rule 1)** *Let $G, P, Q$ be a GSPR instance, and let $G'$ be obtained from $G$ by deleting all vertices and edges that do not lie on an S-path. Then*

- *$G, P, Q$ is again a GSPR instance, and*
- *$P \leadsto_G Q$ if and only if $P \leadsto_{G'} Q$.*

Next, we want to ensure that $G$ contains no cut vertex. In a graph $G$ where every vertex lies on an S-path, it is easily seen that every cut vertex $v$ separates $s$ from $t$, i.e. $s$ and $t$ lie in different components of $G - v$ (so $v \neq s, t$). In addition, $v$ is the only vertex in its layer. This motivates the following definition: if $v$ is a cut vertex of $G$, then *splitting* the instance at $v$ yields two instances $G_{sv}, P_{sv}, Q_{sv}$ and $G_{vt}, P_{vt}, Q_{vt}$, defined as follows: $G_{sv}$ is the subgraph of $G$ induced by all vertices that lie on a shortest $sv$-path, and $G_{vt}$ is the subgraph of $G$ induced by all vertices that lie on a shortest $vt$-path. Let $P_{sv} = P \cap V(G_{sv})$, $Q_{sv} = Q \cap V(G_{sv})$, $P_{vt} = P \cap V(G_{vt})$ and $Q_{vt} = Q \cap V(G_{vt})$. Observe that $V(G_{sv}) \cup V(G_{vt}) = V(G)$, and $V(G_{sv}) \cap V(G_{vt}) = \{v\}$. Thus for every S-path $P'$ in $G$, $P' \cap V(G_{sv})$ is a shortest $sv$-path, and $P' \cap V(G_{vt})$ is a shortest $vt$-path. Using these observations, the following proposition follows easily.

**Proposition 22 (Rule 2)** *Let $G, P, Q$ be a GSPR instance in which every vertex lies on an S-path. If $G$ contains a cut vertex $v$, then*

- *splitting at $v$ yields GSPR instances $G_{sv}, P_{sv}, Q_{sv}$ and $G_{vt}, P_{vt}, Q_{vt}$, and*
- *$P \leadsto_G Q$ if and only if both $P_{sv} \leadsto_{G_{sv}} Q_{sv}$ and $P_{vt} \leadsto_{G_{vt}} Q_{vt}$.*

Finally, we can reduce neighborhood-dominated vertices $z$ simply by deleting them, and adjusting $P$ and $Q$ appropriately. The next proposition is again trivial. (We note however that the analog statement does not hold for TSPR, which explains why Theorem 8 is only formulated for GSPR.)

**Proposition 23 (Rule 3)** *Let $G, P, Q$ be a GSPR instance, and let $z, z'$ be a pair of vertices with $N(z) \subseteq N(z')$ and $z \neq s, t$. Let $G' = G - z$, and let $P'$ and $Q'$ be obtained from $P$ and $Q$ respectively by replacing every occurrence of $z$ by $z'$. Then $P \rightsquigarrow_G Q$ if and only if $P' \rightsquigarrow_{G'} Q'$.*

By applying the above three reduction rules exhaustively to a given GSPR instance $G, P, Q$, we obtain a set of reduced instances in polynomial time, on which it suffices to solve the problem.

**Theorem 8** *Let $G, P, Q$ be a GSPR instance. In polynomial time, a set of reduced GSPR instances can be constructed such that $P \rightsquigarrow_G Q$ if and only if for every reduced instance $G_i, P_i, Q_i$, it holds that $P_i \rightsquigarrow_{G_i} Q_i$. If $G$ is plane, all of the reduced instances are plane. The sum of the number of edges of the reduced instances is at most $|E(G)|$.*

**Proof:** First, delete all vertices and edges of $G$ that do not lie on S-paths (Rule 1). Next, as long as there exists a neighborhood-dominated vertex $z$, delete it and adjust the paths $P$ and $Q$ appropriately (Rule 3). Finally, if there is a cut vertex $v$, split the instance at $v$ (Rule 2). Continue splitting the resulting instances as long as there are cut vertices. The resulting instances are reduced: Applying Rule 2 and 3 maintains the property that all vertices and edges lie on S-paths. Applying Rule 2 maintains the property that there are no neighborhood-dominated vertices.

By Propositions 21–23, the result is a set of GSPR instances that are all YES-instances if and only if the original instance $G, P, Q$ is a YES-instance. We also argue that this reduction is *constructive*, as discussed in Section 2: It is easily seen that an S-path $Q'$ with $Q \subseteq Q'$ and $P \rightsquigarrow_G Q$ is obtained by taking the union of the corresponding S-paths $Q'_i$ for the reduced instances, and subsequently, replacing $z'$ by $z$ for every neighborhood-dominated vertex $z \in Q$ (with $z$ and $z'$ defined as in Proposition 23).

Finally, note that applying these rules exhaustively can be done in polynomial time, that the total number of edges does not increase, and that a possible plane embedding can be maintained. □

### B.3 The proofs of Proposition 9 and Theorem 10

**Proposition 9** *Let $G$ be a 2-connected plane graph in which every vertex and edge lies on a shortest $st$-path. For every face $f$ of $G$, there is exactly one local maximum and one local minimum.*

**Proof:** Let $v_0, v_1, \ldots, v_k, v_0$ be the facial cycle corresponding to $f$. By considering a vertex incident with $f$ with maximum $s$-distance, we see that there is at least one local maximum (as observed above, adjacent vertices cannot have the same $s$-distance). Now suppose that there are at least two local maxima, say w.l.o.g. $v_0$ and $v_\ell$ are local maxima for $f$. So $\ell \geq 2$ and $\ell \leq k - 1$. Let $x$ be a vertex with minimum $s$-distance among $v_1, \ldots, v_{\ell-1}$, and let $y$ be a vertex with minimum $s$-distance among $v_{\ell+1}, \ldots, v_k$. Then $\max\{\text{dist}_s(x), \text{dist}_s(y)\} < \min\{\text{dist}_s(v_0), \text{dist}_s(v_\ell)\}$. By combining a shortest path from $s$ to $x$ and a shortest path from $s$ to $y$, we find a path $P$ in $G$ from $x$ to $y$ in which all $s$-distances are at most $\max\{\text{dist}_s(x), \text{dist}_s(y)\}$. Similarly, by combining shortest paths to $t$, we find a path $P'$ from $v_0$ to $v_\ell$ in which all $s$-distances are at least $\min\{\text{dist}_s(v_0), \text{dist}_s(v_\ell)\}$. Thus $P$ and $P'$ are disjoint. Considering the order of $v_0, x, v_\ell, y$ along the facial cycle of $f$, this contradicts that $G$ is planar. An analog argument shows that there is at most one local minimum. □

With a similar argument, we can prove the following claim, which is illustrated in Figure 6 (a):

**Proposition 24** *Let $G$ be a 2-connected plane graph in which every vertex and edge lies on a shortest $st$-path. For every $i \in \{1, \ldots, d-1\}$ and every $v \in L_i$, the neighbors of $v$ can be numbered $v_0, \ldots, v_{p-1}$ in clockwise order around $v$ such that for some $1 \leq j \leq p-1$, $\{v_0, \ldots, v_{j-1}\} \subseteq L_{i-1}$ and $\{v_j, \ldots, v_{p-1}\} \subseteq L_{i+1}$. Every vertex $v \neq s, t$ of $G$ is a local maximum or local minimum for all incident faces except two.*

**Proof:** Let $v \in L_i$. Call neighbors of $v$ in $L_{i-1}$ *in-neighbors*, and neighbors in $L_{i+1}$ *out-neighbors*. Since every vertex lies on a shortest $st$-path, it is clear that both neighbor sets are non-empty. That both the in-neighbors and the out-neighbors are consecutive in the clockwise order around $v$ follows by the following argument: for any two out-neighbors $o_1$ and $o_2$ of $v$, using shortest paths from $o_1$ and $o_2$ respectively to $t$, we can easily construct a cycle $C^+$ containing $o_1, v, o_2$, in which $v$ is the unique vertex with the lowest $s$-distance. Similarly, for any two in-neighbors $i_1$ and $i_2$ of $v$, we can construct a cycle $C^-$ containing $i_1, v, i_2$, in which $v$ is the unique vertex with the highest $s$-distance. Since the cycle $C^-$ and $C^+$ share only the vertex $v$, it follows that the vertices $o_1, o_2$ and $i_1, i_2$ lie in a non-crossing order around $v$.

Next, since $G$ is 2-connected, there is a bijective correspondence between incident faces and consecutive neighbor pairs. Therefore, the previous statement immediately implies that every vertex $v \neq s, t$ of $G$ is a local maximum or local minimum for all incident faces except two. □

The transformation in Section 4 from a TSPR instance $G, P, Q$ to an RSPR instance $G', P, Q$ is defined in terms of faces. Using Proposition 24, we can see how this transformation changes the neighborhood of every vertex of the graph. The following observations are summarized in Figure 6. Consider a vertex $v \neq s, t$, with $v \in L_i$. Consider a face $f$ of $G$ incident with $v$, and let $w$ and $w'$ the two (consecutive) neighbors of $v$ such that $vw$ and $vw'$ are also incident with $f$. (Recall that $G$ is 2-connected, so there are exactly two such neighbors, and they are consecutive.) Note that in the construction of $G'$, $v$ receives exactly one new neighbor $u$ that is drawn in $f$, and thus $u$ appears between $w$ and $w'$ in the clockwise order around $v$. Furthermore, if $w, w' \in L_{i-1}$ then $u \in L_{i-1}$. If $w, w' \in L_{i+1}$ then $u \in L_{i+1}$. If $w$ and $w'$ are in different layers, then $u \in L_i$, and thus $uv$ is a layer edge. Using Proposition 24, we can now conclude that every $v \neq s, t$ is incident with exactly two layer edges.

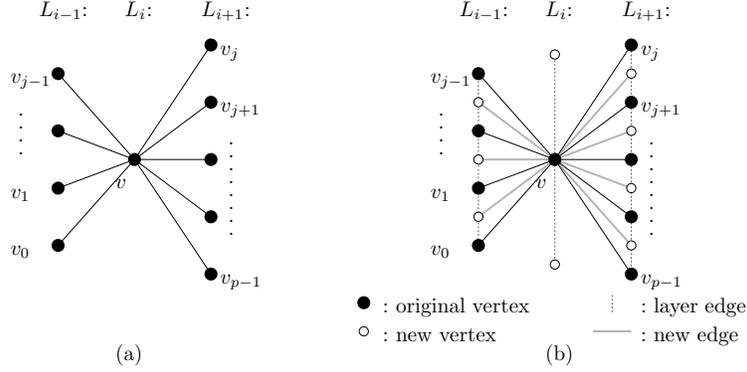

**Fig. 6.** How the transformation changes the neighborhood of a vertex $v \in L_i$.

We will first argue that $G', P, Q$ is an RSPR instance in standard form (Lemma 25). Afterwards, we will show that $P \leadsto_G^T Q$ if and only if $P \leadsto_{G'}^R Q$ (Lemma 26 below). We remark that it may not be immediately obvious that $G'$ is simple: we might have added a layer edge between a vertex pair $a, b$ twice, for different faces. The simplicity of $G'$ is also shown in the next proof.

**Lemma 25** *Let $G, P, Q$ be a reduced TSPR instance, and let $G', P, Q$ be the instance that results from the transformation given in Section 4. Then $G', P, Q$ is in standard form, and $G'$ is simple.*

**Proof:** We prove that the properties (i)–(iii) from Definition 6 hold for $G'$.

(i) We prove that for every $i \in \{1, \ldots, d-1\}$, every vertex in $G'[L_i]$ has degree 2. For every $i$, all edges of $G'[L_i]$ are layer edges. As observed above (see Figure 6), every original vertex is incident with exactly two layer edges. By construction, the same holds for every new vertex.
(ii) We will now prove that for every $i \in \{1, \ldots, d-1\}$ and $v \in L_i$, $G'[L_{i-1} \cap N(v)]$ is a path. If $i = 1$ the statement is clear, so assume $i \geq 2$. First we will show that $G'[L_{i-1}]$ is a cycle.
We showed above that $G'[L_{i-1}]$ is 2-regular, so it is a set of cycles. Consider the set of simple closed curves in the plane corresponding to these cycles. Every S-path contains exactly one vertex from $L_{i-1}$, so the corresponding simple curve from $s$ to $t$ crosses exactly one of these curves. (That it actually crosses the curve, and not just touches it, follows again by considering the local structure around the vertex; see Figure 6.) Hence $G'[L_{i-1}]$ has exactly two faces, one containing $s$ and one containing $t$, and therefore it is in fact a single cycle.

Next, we prove that for every $i$ and $v \in L_i$, $G'[L_{i-1} \cap N(v)]$ is connected. Suppose it is not connected and consider two neighbors $x, y$ in different components of $G'[L_{i-1} \cap N(v)]$. Since $C^1 = G'[L_{i-1}]$ is a cycle, we can choose vertices $x'$ and $y'$ on this cycle that are not adjacent to $v$, and that lie in different components of $C^1 - x - y$. Consider a cycle $C^2$ that is obtained by combining a shortest $sv$-path through $x$ with a shortest $sv$-path through $y$. The curve given by $C^2$ divides the plane into two regions; call the region that contains $t$ that *outside* and the other region the *inside*. Then one of $x'$ and $y'$ lies inside $C^2$, and one lies outside. W.l.o.g. assume that $x'$ lies inside. Consider a shortest path $P$ from $x'$ to $t$, and let $z$ be the vertex in $L_i$ on this path (the first vertex after $x'$). The path $P$ must cross $C^2$. Since $v$ is the only vertex in a layer higher than $i-1$ in $C^2$, it follows that $v = z$, contradicting that $x' \notin N(v)$.

Hence $G[L_{i-1} \cap N(v)]$ is a connected subgraph of the cycle $G[L_{i-1}]$. We conclude the proof by observing that there exists at least one vertex in $L_{i-1}$ that is not adjacent to $v$. Indeed, if $v$ is adjacent to $j$ original vertices in $L_{i-1}$, then it is adjacent to $j-1$ new vertices in $L_{i-1}$ (see Figure 6). But $G'[L_{i-1}]$ is a cycle in which every edge is incident with exactly one new vertex and one original vertex, so it contains an equal number of original vertices and new vertices. So there is at least one new vertex in $L_{i-1}$ that is not adjacent to $v$, and thus $G[L_{i-1} \cap N(v)]$ is a path.

(iii) Finally, we have to show that for any two $u, v \in L_i$ with $uv \in E(G')$, it holds that $|N(u) \cap N(v) \cap L_{i-1}| \leq 1$. This follows easily since $uv$ is a layer edge, so exactly one of $u$ and $v$ is a new vertex, which by construction has only one neighbor in $L_{i-1}$.

Now it follows easily that $G'$ is simple: $G$ is reduced and therefore 2-connected, so every layer $L_i$ in $G$ with $1 \leq i \leq d-1$ contains at least two vertices. As showed above, in $G'$, $L_i$ contains exactly twice as many vertices, and $G[L_i]$ is a cycle. So it is a cycle of length at least 4, and therefore no multi-edges are introduced during the construction of $G'$. □

**Lemma 26** *Let $G, P, Q$ be a reduced TSPR instance, and let $G', P, Q$ be the RSPR instance obtained from it by the transformation given in Section 4. Then $P \leadsto_G^T Q$ if and only if $P \leadsto_{G'}^R Q$.*

**Proof:** Suppose $P \leadsto_G^T Q$, and consider a corresponding topological rerouting sequence in $G$. For any rerouting step $x, a, y \to x, b, y$ in this sequence, $C = x, a, y, b, x$ is a cycle in $G$ that does not separate $s$ from $t$. If there is at least one vertex $z$ that is separated from $s$ and $t$ by $C$, then since every vertex and edge of $G$ lies on an S-path, $z$ is part of the same layer as $a$ and $b$, and $N(z) = \{x, y\}$. But then $N(z) \subseteq N(a)$, which contradicts that $G, P, Q$ is a reduced instance. Therefore, $C$ is a facial cycle. Let $f$ be the corresponding face. Then in $G'$, a new vertex $z$ is added in $f$, with $N(z) = \{x, a, b, y\}$. So in $G'$, the rerouting step $x, a, y \to x, b, y$ can be replaced by two rerouting steps $x, a, y \to x, z, y \to x, b, y$, which are both admissible in $\mathrm{SP}^R(G', s, t)$ (since $az, bz \in E(G')$). Adapting every rerouting step in the sequence this way gives a rerouting sequence from $P$ to $Q$ in $\mathrm{SP}^R(G', s, t)$, so $P \leadsto_{G'}^R Q$.

Next, we prove the converse. Suppose $P \leadsto_{G'}^R Q$, so we may consider a restricted rerouting sequence $Q_0, \ldots, Q_k$ with $Q_0 = P$ and $Q_k = Q$. We will transform this to a topological rerouting sequence from $P$ to $Q$ in $\mathrm{SP}^T(G, s, t)$. This requires the following claim.

*Claim*: If a path $Q_i$ in this sequence contains vertices $w_1, w_2, w_3$ in layers $L_{i-1}$, $L_i$ and $L_{i+1}$ respectively, such that $w_2$ is a new vertex, then $w_1$ and $w_3$ are original vertices, and there exists an original vertex $w_2'$ in $L_i$ that is adjacent to both $w_1$ and $w_3$.

This claim holds because it is not possible that a change $x, a, y \to x, b, y$ is made where $x$ or $y$ is a new vertex, since new vertices have exactly one neighbor in previous and next layers. For this reason, we can map every path $Q_j$ in the sequence to a path $Q_j'$ in $G$: replace every new vertex $w_2$ by an original vertex $w_2'$ with $N(w_2) \subseteq N(w_2')$ (as described in the above claim). To be precise, these paths have to be modified in increasing order of $j$, such that if $w_2 \in Q_j \cap L_i$ is a new vertex, then it should be replaced by the original vertex $w_2' \in Q_{j-1}' \cap L_i$. Observe that then the resulting sequence $Q_0', \ldots, Q_k'$ yields a rerouting sequence for $G$ from $P$ to $Q$ (after removing duplicates). It remains to prove that this is a *topological* rerouting sequence. Since all layer edges are between one new vertex and one original vertex, every rerouting step $x, a, y \to x, b, y$ in the new sequence corresponds to a pair of rerouting steps $x, a, y \to x, z, y \to x, a, y$ in the original sequence, where $z$ is a new vertex that is adjacent to both the original vertices $a$ and $b$. Such a vertex exists only if $x, a, y, b, x$ is a facial 4-cycle in $G$. Hence $x, a, b, y$ is not a switch, and therefore the rerouting sequence from $P$ to $Q$ in $G$ is a topological rerouting sequence. □

Now we can prove Theorem 10.

**Theorem 10** *Let $G, P, Q$ be a reduced TSPR instance. In polynomial time, it can be decided whether $P \leadsto_G^T Q$.*

**Proof:** The transformation that constructs $G', P, Q$ can be done in polynomial time. By Lemma 25, $G', P, Q$ is an RSPR instance in standard form. Lemma 26 shows that $P \leadsto_G^T Q$ if and only if $P \leadsto_{G'}^R Q$. Theorem 7 shows that the latter can be decided in polynomial time. □

## C  Proofs omitted from Section 5

### C.1  The proof of Lemma 11

Recall the following definition.

**Definition 27** *Let $x$, $y$ be a switch pair in $G$, with $x \in L_i$ and $y \in L_{i+2}$ for some $i$. Then $G_{sy}$ is the subgraph of $G$ induced by all vertices that lie on a shortest $sy$-path, and $G_{xt}$ is the subgraph of $G$ induced by all vertices that lie on a shortest $xt$-path. For an S-subpath $Q$, we denote $Q_{xt} = Q \cap V(G_{xt})$, and $Q_{sy} = Q \cap V(G_{sy})$.*

We remark that the rerouting sequences that we consider in $G_{sy}$ ($G_{xt}$), consist of shortest $sy$-paths (resp. $xt$-paths), which are called S-paths in the respective graphs. Let $x \in L_i$. Note that the fact that $x, y$ is a switch pair implies for instance that $L_j \subseteq V(G_{xt})$ holds for all $j \geq i + 2$. In addition, every vertex in $L_{i+1}$ is part of at least one of $G_{sy}$ and $G_{xt}$.

**Definition 28** *Let $x, y$ be a switch pair, and let $P$ be an S-path that contains $x$ and $y$. For any S-path $Q$, we define the vertex set $P_{xt}(Q)$ to firstly contain all vertices of $Q_{xt}$. Secondly, for every $i$ with $Q_{xt} \cap L_i = \emptyset$, it contains the vertex $P \cap L_i$. Similarly, the vertex set $P_{sy}(Q)$ contains all vertices of $Q_{sy}$. In addition, for every $i$ with $Q_{sy} \cap L_i = \emptyset$, it contains the vertex $P \cap L_i$.*

**Proposition 29** *Let $x, y$ be a switch pair and let $P$ be an S-path that contains $x$ and $y$. For any S-path $Q$, both $P_{xt}(Q)$ and $P_{sy}(Q)$ are S-paths.*

**Proof:** Let $x \in L_i$ and $y \in L_{i+2}$. We prove the statement for $P_{xt}(Q)$. By definition, it contains exactly one vertex from every layer. It remains to show that its vertices in subsequent layers are adjacent. Most cases follow easily from the definitions. The only nontrivial case is where $Q \cap L_i \neq \{x\}$, and $Q \cap L_{i+2} \neq \{y\}$. In that case, since $x, a, y, b, x$ is a cycle in $G$ that separates $s$ from $t$, $Q$ must contain $a$ or $b$. Therefore $P_{xt}(Q)$ contains this vertex as well (since $\{a, b\} \subseteq V(G_{xt})$). This shows that the vertices of $P_{xt}(Q)$ in the layers $L_i$, $L_{i+1}$ and $L_{i+2}$ indeed form a path. □

We are now ready to prove the Lemma 11[1].

**Lemma 11** *Let $G, P, Q$ be a plane reduced GSPR instance, such that $x, y$ is a switch pair with $\{x, y\} \subseteq P$, and $Q$ is one of the following:*
*(i) an S-path that contains $x$ and $y$,*
*(ii) $Q = \{x', y'\}$ where $x', y'$ is a switch pair, or*
*(iii) $|Q| = 1$.*
*Then $P \rightsquigarrow_G Q$ if and only if both $P_{sy} \rightsquigarrow_{G_{sy}} Q_{sy}$ and $P_{xt} \rightsquigarrow_{G_{xt}} Q_{xt}$.*

**Proof:** First of all, we observe that $P_{sy}$ is a shortest $sy$-path in $G_{sy}$, and $Q_{sy}$ is a shortest $sy$-subpath. Hence the notation $P_{sy} \rightsquigarrow_{G_{sy}} Q_{sy}$ is well-defined. ($Q_{sy}$ may however be empty, in which case the statement is trivially true.) An analog statement holds for the paths in $G_{xt}$. In the remainder of this proof we will omit the subscripts and simply write e.g. $P \rightsquigarrow Q$.

---
[1] It might be an insightful exercise to verify that the 'if' part of Lemma 11 does not hold for arbitrary S-paths $Q$.

We first prove that $P \rightsquigarrow Q$ implies $P_{sy} \rightsquigarrow Q_{sy}$. Consider a rerouting sequence $Q_0, \ldots, Q_k$ with $Q_0 = P$ and $Q_k = Q$. Then after removing duplicates from the sequence, $P_{sy}(Q_0), \ldots, P_{sy}(Q_k)$ gives a rerouting sequence from $P$ to $P_{sy}(Q)$ (Proposition 29). This in turn gives a rerouting sequence from $P_{sy}$ to $Q_{sy}$ in $G_{sy}$, so $P_{sy} \rightsquigarrow Q_{sy}$. Analogously, it follows that for $G_{xt}$, $P_{xt} \rightsquigarrow Q_{xt}$ holds.

Now suppose that both $P_{sy} \rightsquigarrow Q_{sy}$ and $P_{xt} \rightsquigarrow Q_{xt}$ hold. We prove that $P \rightsquigarrow Q$ follows, by considering three cases.

*Case 1:* Suppose that $Q_{sy} = Q$ or $Q_{xt} = Q$ holds. (This covers in particular the case $|Q| = 1$.) W.l.o.g. assume $Q = Q_{xt}$. Then there exists a rerouting sequence from $P_{xt}$ to a shortest $xt$-path $Q'$ in $G_{xt}$ with $Q \subseteq Q'$. Apply the rerouting steps from this sequence to the shortest $st$-path $P$ in $G$. This results in a shortest $st$-path in $G$ that contains $Q$, which implies $P \rightsquigarrow Q$.

*Case 2:* Next, consider the case where $Q$ is an S-path that contains $x$ and $y$. Similar to Case 1, the rerouting sequence from $P_{xt}$ to $Q_{xt}$ in $G_{xt}$ yields a rerouting sequence from $P$ to $P_{xt}(Q)$ in $G$. Denote $P' = P_{xt}(Q)$. Since $P'$ contains both $x$ and $y$ again, we can subsequently apply rerouting steps from a rerouting sequence from $P_{sy}$ to $Q_{sy}$ in $G_{sy}$ to the path $P'$, which results in the path $P'_{sy}(Q) = Q$ (Here we apply Definition 28 with $P'$ in the role of $P$).

*Case 3:* In the remaining case, $Q = \{x', y'\}$ where $x', y'$ is a switch pair, but neither $Q \subseteq V(G_{sy})$ nor $Q \subseteq V(G_{xt})$ holds. Let $x' \in L_j$ and $y' \in L_{j+2}$, and $x \in L_i$ and $y \in L_{i+2}$. From the fact that $x, a, b, y$ is a switch, it follows that layer $i+2$ and all subsequent layers are included entirely in $G_{xt}$, so $j \leq i+1$. Analogously, $j \geq i-1$ follows. If $j = i+1$, then $x' \notin V(G_{xt})$ implies that $y$ is the unique neighbor of $x'$ in $L_{i+2} = L_{j+1}$ (using again that $x, a, b, y$ is a switch). But this contradicts that $x', y'$ is a switch pair. By excluding the case $j = i-1$ analogously, it follows that $j = i$. So $x$ and $x'$ are part of the same layer but distinct, and the same holds for $y$ and $y'$. The cycle $x, a, y, b, x$ separates $s$ from $t$, so the only possible common neighbors of $x'$ and $y'$ are $a$ and $b$. Since $x', y'$ is a switch pair, they have at least two common neighbors, so we conclude that $N(x') \cap N(y') = \{a, b\}$. In other words, $x', a, b, y'$ is the switch that makes $x', y'$ a switch pair. This implies that for every $z \in L_{i+1} \setminus \{a, b\}$, either $N(z) = \{x, y'\}$ or $N(z) = \{x', y\}$. Both are strict subsets of $N(a)$, so since $G$ is reduced, such a vertex $z$ does not exist. We conclude that $L_{i+1} = \{a, b\}$.

Using this information, we can conclude the proof analogously to the previous cases: $P_{xt} \rightsquigarrow Q_{xt}$ with $Q_{xt} = \{y'\}$ implies that there exists a rerouting sequence in $G_{xt}$ from $P_{xt}$ to a shortest $xt$-path that contains $y'$. Apply the rerouting steps from this rerouting sequence to $P$, to obtain a rerouting sequence that ends with a shortest $st$-path $Q'$ in $G$ that contains both $x$ and $y'$. Subsequently, consider a rerouting sequence from $P_{sy}$ to a path in $G_{sy}$ that contains $x'$. Every path in this sequence contains $y$. Since $N(y) \cap L_{i+1} = \{a, b\} = N(y') \cap L_{i+1}$, we can replace $y$ by $y'$ in every path in the sequence, which gives a sequence of shortest $sy'$-paths in $G$. The first path of this sequence coincides with (is a subset of) $Q'$. Therefore, applying these rerouting steps to $Q'$ ends with a shortest $st$-path in $G$ that contains both $x'$ and $y'$. We conclude that $P \rightsquigarrow Q$. □

## C.2 The proof of Theorem 13

**Theorem 13** *Let $G$ be a plane graph, and let $P$ and $Q$ be S-paths in $G$. In polynomial time, it can be decided whether $P \rightsquigarrow Q$.*

**Proof:** By Theorem 8, we may assume that the instance is reduced. First, we decide whether $P \rightsquigarrow^T Q$, which can be done in polynomial time (Theorem 10). If not, then for every switch pair $x, y$, we decide whether $P \rightsquigarrow \{x, y\}$ and $Q \rightsquigarrow \{x, y\}$. By Theorem 12, and using the fact that the number of switches is polynomial, this can be done in polynomial time. Furthermore, if the answer is affirmative, then corresponding paths containing $x$ and $y$ are computed, that are reachable from $P$ and $Q$ respectively. Clearly, if there exists a switch pair $x, y$ such that $P \rightsquigarrow \{x, y\}$ but $Q \not\rightsquigarrow \{x, y\}$ or vice versa, then $P \not\rightsquigarrow Q$, so we may answer negatively. So now assume that the same set of switch pairs is reachable from both $P$ and $Q$. If no switch pair is reachable, then we may conclude that $P \not\rightsquigarrow Q$. (Indeed, if $P \rightsquigarrow Q$, then either this is a topological rerouting sequence, or it contains a rerouting step $x, a, y \to x, b, y$ where $x, a, b, y$ is a switch. When considering the first such rerouting step, it holds that $P \rightsquigarrow^T \{x, y\}$, and thus $P \rightsquigarrow \{x, y\}$.)

In the remaining case, there exists a switch pair $x, y$ that is reachable from both $P$ and $Q$. Furthermore, corresponding S-paths $P'$ and $Q'$ have been computed with $P \rightsquigarrow P'$, $\{x, y\} \subseteq P'$, $Q \rightsquigarrow Q'$ and $\{x, y\} \subseteq Q'$. Note that $P \rightsquigarrow Q$ if and only if $P' \rightsquigarrow Q'$. Then we reduce to the SPR instances $G_{sy}, P'_{sy}, Q'_{sy}$ and $G_{xt}, P'_{xt}, Q'_{xt}$. By Lemma 11, it now suffices to decide whether $P'_{sy} \rightsquigarrow_{G_{sy}} Q'_{sy}$ and $P'_{xt} \rightsquigarrow_{G_{xt}} Q'_{xt}$. This is done recursively.

It remains to consider the complexity of this algorithm. We argued that the complexity of the above procedure, not counting the recursive calls, can be bounded by a (monotone increasing) polynomial $\text{poly}(n)$, where $n = |V(G)|$. Analogously to the proof of Theorem 12, an induction proof over the distance $d$ between the end vertices $s$ and $t$ now shows that that the entire algorithm terminates in time $(\frac{4}{3}d - 3) \cdot \text{poly}(n)$ if $d \geq 3$, and $\text{poly}(n)$ if $d \leq 2$. (Of course the function $\text{poly}(n)$ is different in the current proof; its degree is higher.) $\square$

## D  A polynomial complexity bound for low degree instances

In this section, we assume that $G, P, Q$ is an instance of RSPR in which every vertex $v \neq s, t$ has at most two neighbors in the previous layer, and at most two neighbors in the next layer. We call such an instance a *low degree instance*. We will show that for low degree instances, the dynamic programming procedure from Section 3 terminates in polynomial time, by showing that for every $i$, the encoding $H_i$ of $G_i, P, Q$ satisfies $|V(H_i)| \leq i \cdot |L_i|$. The proof is similar to the proof of Theorem 7.

Define a layer edge $uv$ with $u, v \in L_i$ to be *useless* if $N(u) \cap N(v) \cap L_{i-1} = \emptyset$. The reason for this is that clearly, no rerouting step in a (restricted) rerouting sequence will change $u$ to $v$ or $v$ to $u$. So in this case, for $G' = G - uv$ it holds that $\mathrm{SP}^R(G, s, t) = \mathrm{SP}^R(G', s, t)$. Hence it suffices to prove the statement for low degree instances without useless edges, which is what we will assume for $G$ throughout this section.

**Proposition 30** *Let $G, P, Q$ be a low degree RSPR instance without useless edges, with distance $d$ from $s$ to $t$. Then for every $i \in \{2, \ldots, d-1\}$, $G[L_i]$ has maximum degree at most 2.*

**Proof:** Consider $u \in L_i$. The vertex $u$ has at most two neighbors in $L_{i-1}$. Each neighbor of $u$ in $L_{i-1}$ has at most one other neighbor in $L_{i+1}$, so it follows that there are at most two vertices $v, w \in L_{i+1}$ that share a neighbor with $u$. □

Throughout, denote by $H_i$ the encoding of $G_i, P, Q$ (Definition 1). By Proposition 16, for every $i$, $l$ is a homomorphism from $H_i$ to $G[L_i]$. Our first goal is to prove that in the case of low degree graphs, this homomorphism is locally injective.

**Lemma 31** *Let $G, P, Q$ be a low degree RSPR instance without useless edges, with distance $d$ from $s$ to $t$. For $i \in \{0, \ldots, d-1\}$, let $H_i$ be the encoding of $G_i, P, Q$. Then for every $i$, the node labels $l$ form a LIH from $H_i$ to $G[L_i]$.*

**Proof:** We prove the statement by induction over $i$. For $i = 0$, the statement is trivial. For $i = 1$, $G_i$ is a $K_k$ with $k = |L_1|$, so $H_i$ has $k$ nodes, and $l$ is a bijection to $L_i$. This proves local injectivity.

Now for the induction step, suppose $l$ is a LIH from $H_i$ to $G[L_i]$, with $i \geq 1$. We prove that $l$ is a LIH from $H_{i+1}$ to $G[L_{i+1}]$.

Consider $a \in V(H_{i+1})$, and the corresponding component $C_a$ of $H_i^v$, where $v = l(a)$. Since $v$ has at most two neighbors in $L_i$, nodes in $C_a$ have at most two different labels $l$ (because these labels are chosen from $N(v) \cap L_i$). Since $C_a$ is connected and by the induction assumption, no node in $H_i$ has different neighbors with the same label $l$, it follows that $|V(C_a)| \leq 2$, and all nodes $x \in V(C_a)$ have different labels $l(x)$.

Using this fact, we can prove that $l$ is a LIH from $H_{i+1}$ to $G[L_{i+1}]$: Suppose not, so there exists a node $a \in V(H_{i+1})$ with neighbors $b, c \in V(H_{i+1})$ such that $l(b) = l(c)$. Let $w = l(b) = l(c)$. This means that there exist nodes $x \in V(C_b) \cap V(C_a)$ and $y \in V(C_c) \cap V(C_a)$ (Proposition 16). But above we argued that $C_a$ consists of a single node, or two adjacent nodes. Hence $x = y$, or $xy \in E(H_i)$. This contradicts that $C_b$ and $C_c$ are distinct components of the *induced* subgraph $H_i^w$. □

**Corollary 32** *Let $G, P, Q$ be a low degree RSPR instance without useless edges, with distance $d$ from $s$ to $t$. For $i \in \{2, \ldots, d-1\}$, the encoding $H_i$ of $G_i, P, Q$ is a path or a cycle.*

**Proof:** For $i \geq 2$, every layer $G[L_i]$ has maximum degree at most 2 (Proposition 30). For every $i$, $l$ is a LIH from $H_i$ to $G[L_i]$ (Lemma 31). So $H_i$ has maximum degree at most 2 as well. By definition, encodings are connected, so $H_i$ is a path or a cycle. □

Recall that for $v \in L_i$, we denote by $n_i(v)$ the number of nodes $x \in V(H_i)$ with $l(x) = v$. Define
$$\text{MAX}_i := \max_{v \in L_i} n_i(v).$$
Using the previous lemma and corollary, we can show that $\text{MAX}_i$ grows at most linearly with $i$, which is done in the next lemma. From this it will follow that for every $i$, the size of $H_i$ is polynomially bounded.

**Lemma 33** *Let $G, P, Q$ be a low degree RSPR instance without useless edges, with distance $d$ from $s$ to $t$. For $i \in \{1, \ldots, d-2\}$, $\text{MAX}_{i+1} \leq \text{MAX}_i + 1$.*

**Proof:** We will prove the statement by proving for an arbitrary vertex $u \in L_{i+1}$ that $n_{i+1}(u) \leq \text{MAX}_i + 1$ holds. Recall that all nodes $a$ of $H_{i+1}$ with $l(a) = u$ correspond to components of $H_i^u$ (Proposition 15), although since $H_{i+1}$ should be connected, not every component of $H_i^u$ corresponds to a node.

We first handle two easy cases. Suppose $i = 1$. Clearly $\text{MAX}_1 = 1$. Since $u \in L_2$ has at most two neighbors in $L_1$, $H_1^u$ has at most two components, which proves that $n_2(u) \leq 2 = \text{MAX}_1 + 1$. So now we may assume that $i \geq 2$. If $u$ has only one neighbor $v$ in $L_i$, then clearly the number of nodes in $H_{i+1}$ with label $u$ is at most the number of nodes with label $v$ in $H_i$, so $n_{i+1}(u) \leq \text{MAX}_i$. So now we may assume that $N(u) \cap L_i = \{v, w\}$.

*Case 1:* Suppose $vw \notin E(G)$.

In this case, nodes of $H_i$ with label $v$ and $w$ cannot be adjacent either (Proposition 16), so every component of $H_i^u$ consists of a single node. Consider $a \in V(H_{i+1})$ with $l(a) = u$, and suppose that the single node $x$ in $C_a$ has label $l(x) = v$. Since $v$ has at most one other neighbor $u'$ in $L_{i+1}$, $a$ has at most one neighbor $b$ in $H_i$, which would have $l(b) = u'$ (Proposition 16). The same argument applies if the single node in $C_a$ has label $w$. So we conclude that all nodes $a \in V(H_{i+1})$ with $l(a) = u$ have degree 1. Since $H_{i+1}$ is a path or a cycle (Corollary 32), there can be at most two such nodes, which is at most $\text{MAX}_i + 1$, so this proves the claim in this case.

*Case 2:* Suppose $vw \in E(G)$.

Consider the component $C$ of $G[L_i]$ that contains $v$ and $w$. By Proposition 30, this component has maximum degree at most 2 (since $i \geq 2$), hence it is a path or a cycle. Number the vertices of $C$ as $v_0, \ldots, v_p$ such that $v = v_j$ and $w = v_{j+1}$ for some $j$, and such that $v_0, \ldots, v_p$ is a path. (In case $C$ is a cycle, $v_0 v_p \in E(G)$.) Since $H_i$ is a path or a cycle (Corollary 32), we may similarly number its nodes $x_0, \ldots, x_q$, such that this sequence is a path. (If $H_i$ is a cycle, then $x_q x_0 \in E(H_i)$.) To bound the number of nodes $a \in V(H_{i+1})$ with $l(a) = u$, we argue that every node in $H_i$ with label $w$ is adjacent to a node with label $v$, except possibly one. Consider a node $x \in V(H_i)$ with $l(x) = w$, and suppose it has no neighbor with label $v$. Since $l$ is a LIH from $H_i$ to $G[L_i]$, and $w = v_{j+1}$ has at most one neighbor in $G[L_i]$ other than $v = v_j$ (which would be $v_{j+2 \bmod (p+1)}$), this implies that $x$ has degree at most one. Hence w.l.o.g. it follows that $x = x_0$ ($x_0$ and $x_q$ are the only vertices in $H_i$ that can have degree less than 2). Furthermore, from the fact that $l$ is a LIH we can conclude that $l(x_0) = v_{j+1}$, $l(x_1) = v_{j+2 \bmod (p+1)}$, etc. Continuing this way, we see that even if the other end node $x_q$ of the path has again label $l(x_q) = v_{j+1} = w$, then $l(x_{q-1}) = v_j = v$. We conclude that there is at most one node in $H_i$ with label $w$ without a neighbor with label $v$. Therefore, the subgraph $H_i^u$, which is induced by the set of nodes with label $v$ and $w$, has at most one component that contains no node with label $v$. We conclude that it has at most $n_i(v)+1$ components, which proves that $n_{i+1}(u) \leq n_i(v)+1 \leq \text{MAX}_i+1$. $\square$

**Theorem 34** *Let $G, P, Q$ be a low degree RSPR instance. Then in polynomial time, it can be decided whether $P \leadsto^R Q$.*

**Proof:** Let $n = |V(G)|$. As observed in the beginning of this section, we may assume that $G$ contains no useless edges. We apply the dynamic programming algorithm from Section 3 to this instance to decide whether $P \leadsto^R Q$. This algorithm computes encodings $H_i$ of $G_i, P, Q$ for every $i$. A simple induction proof based on Lemma 33 shows that for every $i$, $\text{MAX}_i \leq i$, so $H_i$ contains at most $i$ nodes with the same label. Therefore, it contains at most $i \cdot |L_i| \in O(n^2)$ nodes in total. So the algorithm terminates in time $n^{O(1)}$ (Theorem 5). $\square$

For the special case of GSPR (see Section 2), the above theorem implies the following result.

**Corollary 35** *Let $G, P, Q$ be a GSPR instance, such that every vertex that lies on an S-path has at most two neighbors in the previous layer, and at most two neighbors in the next layer. Then in polynomial time, it can be decided whether $P \leadsto Q$.*